\documentclass[
    pdflatex,
    sn-chicago,
]{sn-jnl}

\jyear{2021}%

\usepackage{bm}
\usepackage{amsmath}
\usepackage{amssymb}
\usepackage{color}

\usepackage[normalem]{ulem}		

\usepackage{lineno}
\newcommand*\patchAmsMathEnvironmentForLineno[1]{%
  \expandafter\let\csname old#1\expandafter\endcsname\csname #1\endcsname
  \expandafter\let\csname oldend#1\expandafter\endcsname\csname end#1\endcsname
  \renewenvironment{#1}%
     {\linenomath\csname old#1\endcsname}%
     {\csname oldend#1\endcsname\endlinenomath}}%
\newcommand*\patchBothAmsMathEnvironmentsForLineno[1]{%
  \patchAmsMathEnvironmentForLineno{#1}%
  \patchAmsMathEnvironmentForLineno{#1*}}%
\AtBeginDocument{%
\patchBothAmsMathEnvironmentsForLineno{equation}%
\patchBothAmsMathEnvironmentsForLineno{align}%
\patchBothAmsMathEnvironmentsForLineno{flalign}%
\patchBothAmsMathEnvironmentsForLineno{alignat}%
\patchBothAmsMathEnvironmentsForLineno{gather}%
\patchBothAmsMathEnvironmentsForLineno{multline}%
}

\usepackage[T1]{fontenc}
\usepackage{newpxtext, mathptmx}

\renewcommand{\eqref}[1]{Eq.~(\ref{#1})}
\newcommand{\figref}[1]{Fig.~\ref{#1}}

\newcommand{\Alfven}{Alfv\'en }
\newcommand{\Alfvenic}{Alfv\'enic }
\newcommand{\MAHTF}{M_{\textrm{A}}^{\textrm{HTF}}}
\newcommand{\MAc}{M_{\textrm{A}}^{\textrm{*}}}
\newcommand{\MAwc}{M_{\textrm{A}}^{\textrm{w*}}}
\newcommand{\MA}{M_{\textrm{A}}}
\newcommand{\MS}{M_{\textrm{S}}}
\newcommand{\VA}{V_{\textrm{A}}}
\newcommand{\thetabn}{\theta_{Bn}}

\newcommand{\wcs}{\Omega_{cs}}
\newcommand{\wci}{\Omega_{ci}}
\newcommand{\wce}{\Omega_{ce}}
\newcommand{\wps}{\omega_{ps}}
\newcommand{\wpi}{\omega_{pi}}
\newcommand{\wpe}{\omega_{pe}}

\newcommand{\betai}{\beta_{i}}
\newcommand{\betae}{\beta_{e}}
\newcommand{\Einif}{E_{i}^{\textrm{NIF}}}
\newcommand{\Eenif}{E_{e}^{\textrm{NIF}}}
\newcommand{\Eehtf}{E_{e}^{\textrm{HTF}}}

\newcommand{\ldiff}{l_{\textrm{diff}}}
\newcommand{\Us}{U_{\textrm{s}}}
\newcommand{\Vs}{V_{\textrm{s}}}
\newcommand{\Ls}{L_{\textrm{s}}}
\newcommand{\Lfeb}{L_{\textrm{FEB}}}

\newcommand{\degr}{^{\circ}}
\newcommand{\phinif}{\phi^{\textrm{NIF}}}
\newcommand{\phihtf}{\phi^{\textrm{HTF}}}
\newcommand{\Bmax}{B_{\textrm{m}}}
\newcommand{\Dmax}{D_{\mu\mu,{\rm max}}}
\newcommand{\gammax}{\gamma_\textrm{max}}

\begin{document}

\title
[Electron Acceleration at Collisionless Shocks]
{Nonthermal Electron Acceleration at Collisionless Quasi-perpendicular Shocks}


\author*[1]{\fnm{Takanobu} \sur{Amano}}\email{amano@eps.s.u-tokyo.ac.jp}

\author[2]{\fnm{Yosuke} \sur{Matsumoto}}

\author[3]{\fnm{Artem} \sur{Bohdan}}

\author[4]{\fnm{Oleh} \sur{Kobzar}}

\author[5]{\fnm{Shuichi} \sur{Matsukiyo}}

\author[6]{\fnm{Mitsuo} \sur{Oka}}

\author[7]{\fnm{Jacek} \sur{Niemiec}}

\author[3,8]{\fnm{Martin} \sur{Pohl}}

\author[1]{\fnm{Masahiro} \sur{Hoshino}}

\affil[1]{
    \orgdiv{Department of Earth and Planetary Science},
    \orgname{The University of Tokyo},
    \orgaddress{
        \city{Tokyo},
        \postcode{113-0033},
        \country{Japan}
    }}

\affil[2]{
    \orgdiv{Department of Physics},
    \orgname{Chiba University},
    \orgaddress{
        \city{Chiba},
        \postcode{263-8522},
        \country{Japan}
    }}

\affil[3]{
    \orgname{Deutsches Elektronen-Synchrotron (DESY)},
    \orgaddress{
        \city{Zeuthen},
        \postcode{D-15738},
        \country{Germany}
    }}


\affil[4]{
    \orgname{Faculty of Materials Engineering and Physics, Cracow University of Technology},
    \orgaddress{
        \city{Krak{\' o}w},
        \postcode{PL-30084},
        \country{Poland}
    }}

\affil[5]{
    \orgdiv{Faculty of Engineering Sciences},
    \orgname{Kyushu University},
    \orgaddress{
        \city{Fukuoka},
        \postcode{816-8580},
        \country{Japan}
    }}

\affil[6]{
    \orgdiv{Space Sciences Laboratory},
    \orgname{University of California Berkeley},
    \orgaddress{
        \city{Berkeley},
        \postcode{94720-7450},
        \country{USA}
    }}

\affil[7]{
    \orgdiv{Institute of Nuclear Physics},
    \orgname{Polish Academy of Sciences},
    \orgaddress{
        \city{Krak{\' o}w},
        \postcode{PL-31342},
        \country{Poland}
    }}

\affil[8]{
    \orgdiv{Institute of Physics and Astronomy},
    \orgname{University of Potsdam},
    \orgaddress{
        \city{Potsdam-Golm},
        \postcode{D-14476},
        \country{Germany}
    }}

\abstract{
Shock waves propagating in collisionless heliospheric and astrophysical plasmas have been studied extensively over the decades. One prime motivation is to understand the nonthermal particle acceleration at shocks. Although the theory of diffusive shock acceleration (DSA) has long been the standard for cosmic-ray acceleration at shocks, plasma physical understanding of particle acceleration remains elusive. In this review, we discuss nonthermal electron acceleration mechanisms at quasi-perpendicular shocks, for which substantial progress has been made in recent years. The discussion presented in this review is restricted to the following three specific topics. The first is stochastic shock drift acceleration (SSDA), which is a relatively new mechanism for electron injection into DSA. The basic mechanism, related in-situ observations and kinetic simulations results, and how it is connected with DSA will be discussed. Second, we discuss shock surfing acceleration (SSA) at very high Mach number shocks relevant to young supernova remnants (SNRs). While the original proposal under the one-dimensional assumption is unrealistic, SSA has now been proven efficient by a fully three-dimensional kinetic simulation. We discuss the multidimensional nature of SSA and its role in electron injection. Finally, we discuss the current understanding of the magnetized Weibel-dominated shock. It is essentially a magnetized shock in which the reflected-gyrating ions dominate the formation of the shock structure but with a substantial magnetic field amplification by the ion-Weibel instability. Spontaneous magnetic reconnection of self-generated current sheets within the shock structure is an interesting consequence of Weibel-generated strong magnetic turbulence. Although the exact condition for active magnetic reconnection has not been clarified, we argue that high Mach number shocks with both \Alfven and sound Mach numbers exceeding 20-40 will likely behave as a Weibel-dominated shock. Despite a number of interesting recent findings, the relative roles of SSDA, SSA, and magnetic reconnection for electron acceleration at collisionless shocks and how the dominant particle acceleration mechanisms change depending on shock parameters remain to be answered.
}




\maketitle

\clearpage
\section{Introduction} \label{sec:intro}

The dilute and high-temperature plasmas in space are often in the collisionless state in which binary Coulomb collisions are negligible for the energy and momentum transfer between the particles. Instead, plasma waves of various spatiotemporal scales are believed to play the role via collective wave-particle interactions. A magnetohydrodynamic (MHD) shock wave must dissipate the supersonic flow kinetic energy in the upstream low-entropy plasma to provide the thermal pressure in the subsonic downstream flow that sustains the stable shock structure. Therefore, the need for the dissipation at the shock indicates that the collisionless shock transition layer will be filled with intense and nonlinear plasma waves that heat the plasma incoming into the shock. Understanding the dynamics of plasma waves has indeed been one of the primary motivations to study collisionless shocks \citep[e.g.,][]{Wu1984a}.

The acceleration of high-energy charged particles around collisionless shocks has attracted even greater attention. The standard paradigm for the origin of cosmic rays (CRs) at present is that shock waves generated by supernova explosions accelerate CRs at least up to the so-called knee energy $10^{15.5}$ eV. The early conjecture based on the energetics argument \citep{Baade1934,Ginzburg1961} has now been supported by the theory of diffusive shock acceleration (DSA) as the mechanism of particle acceleration \citep{Bell1978a,Blandford1978a,Drury1983,Blandford1987}. Observations of radio and X-ray synchrotron emissions from young supernova remnants (SNRs) clearly demonstrate the presence of ultra-relativistic electrons \citep[see, ][and references therein]{Reynolds2008}. More recently, observations of hadronic $\gamma$-rays confirm that protons, the primary population of CRs, are also accelerated to relativistic energies \citep[e.g.,][]{Abdo2010ApJ_IC443,Abdo2010Science_W44,Ackermann2013_Science}. It is now well established that strong astrophysical collisionless shocks can produce relativistic electrons and protons. However, it is not yet known under what conditions the accelerated particle energies reach PeV.

The collisionless shocks have most extensively been studied with in-situ spacecraft observations in the heliosphere. Although the scale sizes are substantially different, direct measurements of both particles and electromagnetic fields along the spacecraft trajectory provide far more information available than with remote-sensing astrophysical observations. Indeed, the basic idea of DSA theory was proven by spacecraft measurements of energetic ions around Earth's bow shock as well as interplanetary shocks driven by coronal mass ejections \citep[e.g.,][]{Lee1982a,Lee1983a,Kennel1986}. However, it remains a puzzle as to why these shocks are very poor in terms of electron acceleration efficiency. This fact is in marked contrast to SNR shocks, where electron acceleration to relativistic energies appears to be more common. Given that plasma waves mediate the energy and momentum transfer in collisionless plasmas, it is natural to anticipate that the property of plasma waves changes depending on the shock parameters, in particular, \Alfven and sound Mach numbers. Note that Mach numbers of typical young SNR shocks ($\sim$ 10--100), particularly during the early phase of evolution, may be much larger than the heliospheric shocks ($\sim$ 2--10).

Although the dynamics of collisionless shock involve broad topics of vastly different scale sizes, this review will focus on the dynamics of electrons and, in particular, the acceleration of nonthermal electrons. It has been challenging to investigate the small-scale electron dynamics both in simulations and observations. However, the situation has been improved dramatically in the last several years. On the one hand, a substantial advance in the spatiotemporal resolution of particle measurement provided by NASA's Magnetospheric Multiscale (MMS) mission contributes significantly to our understanding the electron-scale physics \citep{Burch2016a}. On the other hand, increasing computational resources with modern supercomputers are now allowing us to perform multidimensional fully kinetic Particle-in-Cell (PIC) simulations with more and more realistic choices of parameters \citep[e.g.,][]{Pohl2020}. It is thus timely to summarize recent findings on the electron dynamics at shocks. We will place a particular emphasis on the dependence on Mach numbers. Understanding the Mach number dependence is crucial in elucidating the discrepancy in the electron acceleration efficiency at shocks in different environments. In this review, we discuss almost exclusively the quasi-perpendicular shock where the upstream magnetic field is inclined by more than $45 \degr$ with respect to the shock normal vector simply because it is better understood than the more complicated quasi-parallel shock. Nevertheless, the knowledge discussed here will also be helpful for understanding the quasi-parallel shock dynamics because large-amplitude transverse fluctuations generated ahead of the shock often make portions of the shock locally like quasi-perpendicular. It should be noted that the topics covered in this review are mainly based on the results obtained from the authors' own work in the last several years. More comprehensive discussions of collisionless shock physics may be found elsewhere \citep[e.g.,][]{Balogh2013a,BurgessScholer2015}.

The rest of the paper will be organized as follows. In Section \ref{sec:2}, we will review classical understanding of the collisionless shock, including dissipative and dispersive fluid descriptions, ion reflection, electron heating. We then introduce the electron injection problem and discuss the reason why the electron acceleration at shocks has been considered a challenging problem. We will then discuss a novel electron injection mechanism called stochastic shock drift acceleration (SSDA) in Section \ref{sec:3}. The classical shock drift acceleration (SDA) will be reviewed and then extended to SSDA by adding stochastic pitch-angle scattering. Observational evidence and relevant kinetic simulations results will be briefly discussed. It will be shown that SSDA under the diffusion approximation can be combined with the standard DSA. In other words, the combined model allows us to discuss both the injection by SSDA and subsequent particle acceleration by DSA on equal footing. The theory indicates that an efficient electron injection will occur preferentially at shocks with higher Mach numbers in the de Hoffmann-Teller frame (HTF). In Section \ref{sec:4}, we discuss shock surfing acceleration (SSA) of electrons as another possible mechanism for electron injection. We note that since SSA requires intense electrostatic waves generated by the electrostatic Buneman instability (BI), it inherently favors shocks with very high Mach numbers. While recent two- and three-dimensional (2D and 3D) simulations indicate that SSA by itself may not be as efficient as originally thought based on earlier one-dimensional (1D) simulations, we suggest that it may still be important as a pre-acceleration mechanism prior to SSDA. In Section \ref{sec:5}, we will continue the discussion on high Mach number shocks but focus on electromagnetic turbulence driven by the Weibel instability (WI). We argue that the well-known shock-surface rippling mode at moderate Mach numbers relevant to Earth's bow shock transitions into the more violent Weibel mode at high Mach numbers. As a result, substantial magnetic field amplification occurs within the shock transition layer. One of the important consequences is spontaneous magnetic reconnection that dissipates the amplified magnetic energy. A fraction of electrons may be accelerated to high energies associated with magnetic reconnection in a highly turbulent shock transition layer. Finally, Section 6 summarizes the paper and presents a brief discussion on the prospect. Note that we have tried to make each section as independent as possible so that readers may skip from one section to another as they wish.


\clearpage
\section{Classical Picture} \label{sec:2}
\subsection{Shock Parameters and Frames} \label{sec:2-parameters-frames}
Before starting the discussion, we introduce important macroscopic parameters that characterize MHD shocks. In this paper, we will discuss only the fast-mode MHD shock, for which the upstream flow speed is greater than the fast magnetosonic wave speed. It might thus appear to be reasonable to investigate the dependence of the shock property on the fast magnetosonic Mach number. However, the property of collisionless shocks is not necessarily regulated by the single Mach number, and both sound and \Alfven Mach numbers play different roles. In addition, the shock propagation angle with respect to the ambient magnetic field orientation is known to  affect the collisionless shock dynamics substantially. We thus mainly discuss the dependence on \Alfven Mach number $\MA$, the plasma beta $\beta = P / (B^2/8\pi)$, and the magnetic obliquity angle $\thetabn$ with respect to the shock normal defined in the far upstream of the shock. The sound Mach number is given by $\MS = \MA \sqrt{2 / \gamma \beta}$, where $\gamma$ is the polytropic index. If individual plasma betas (or temperatures) of ion and electrons become important, $\beta_{s} = P_{s} / (B^2/8\pi)$ will be used with $s = i, e$ indicating ions and electrons, respectively. As we have introduced already, shocks with $\thetabn \gtrsim 45 \degr$ will be referred to as quasi-perpendicular shocks and otherwise quasi-parallel shocks. It is well known that these different types of shocks have substantially different appearances.

We define the Mach numbers in the conventional normal incidence frame (NIF), in which the shock is at rest and the upstream flow is parallel to the shock normal. This is the frame that has been used in almost all numerical simulations. Another important frame that is convenient for theoretical analysis is the so-called de Hoffmann-Teller frame (HTF). In general, HTF is defined as the frame in which the plasma flow is everywhere aligned to the local magnetic field direction. This indicates that the motional electric field vanishes in this special frame, which makes it easy to analyze charged particle motions. One can find HTF by Lorentz transformation from NIF in the direction transverse to the shock normal. However, since the frame transformation velocity should not exceed the speed of light, HTF is non-existent at a shock with $\thetabn$ very close to $90\degr$. A subluminal (superluminal) shock is defined as an oblique shock for which one can (cannot) find HTF. It is important to note that, at a superluminal shock, even a particle traveling along the local magnetic field with the speed of light cannot escape upstream away from the shock. This thus imposes a substantial constraint on particle acceleration at a superluminal shock \citep[e.g.,][]{Takamoto2015b}. Note that non-relativistic oblique shocks of our interest are practically always subluminal because, for instance, $\thetabn > 89\degr$ is needed for typical high-speed young SNR shocks to become superluminal. More detailed discussion on the shock frames can be found in \citep{BurgessScholer2015}.

We will use the standard notations such as the speed of light $c$, the electromagnetic field $\bm{E}, \bm{B}$, the plasma and cyclotron frequency for each species $\wps$, $\wcs$ throughout the paper. The physical quantities in the far upstream will be denoted by the subscript $_{0}$, e.g., $n_0$, $B_0$. We will denote the flow speed in NIF by $\bm{U}$, while that in HTF by $\bm{V}$. The shock speed in NIF and HTF will be denoted by $\Us$ and $\Vs = \Us/\cos \thetabn$, respectively. Later we will find it useful to define the \Alfven Mach number in HTF by $\MAHTF = \MA/\cos \thetabn$.

\subsection{Fluid Description} \label{sec:2-fluid}

Let us first discuss the shock property with the fluid description. The high downstream plasma pressure that balances with the upstream dynamic pressure requires irreversible heating across the shock. Historically, the existence and internal structure of collisionless shocks were first investigated using a dissipative MHD model, which explicitly considers the shock heating with phenomenological dissipation coefficients such as the resistivity, viscosity, and thermal conductivity \citep{Coroniti1970}. However, in collisionless plasmas, these transport coefficients cannot be written in simple forms as they are determined as a result of collective wave-particle interactions. It is, nevertheless, possible to discuss what kinds of dissipation are needed for the stable shock structure.

\begin{figure}[tbp]
    \centering
    \includegraphics[width=1.00\textwidth]{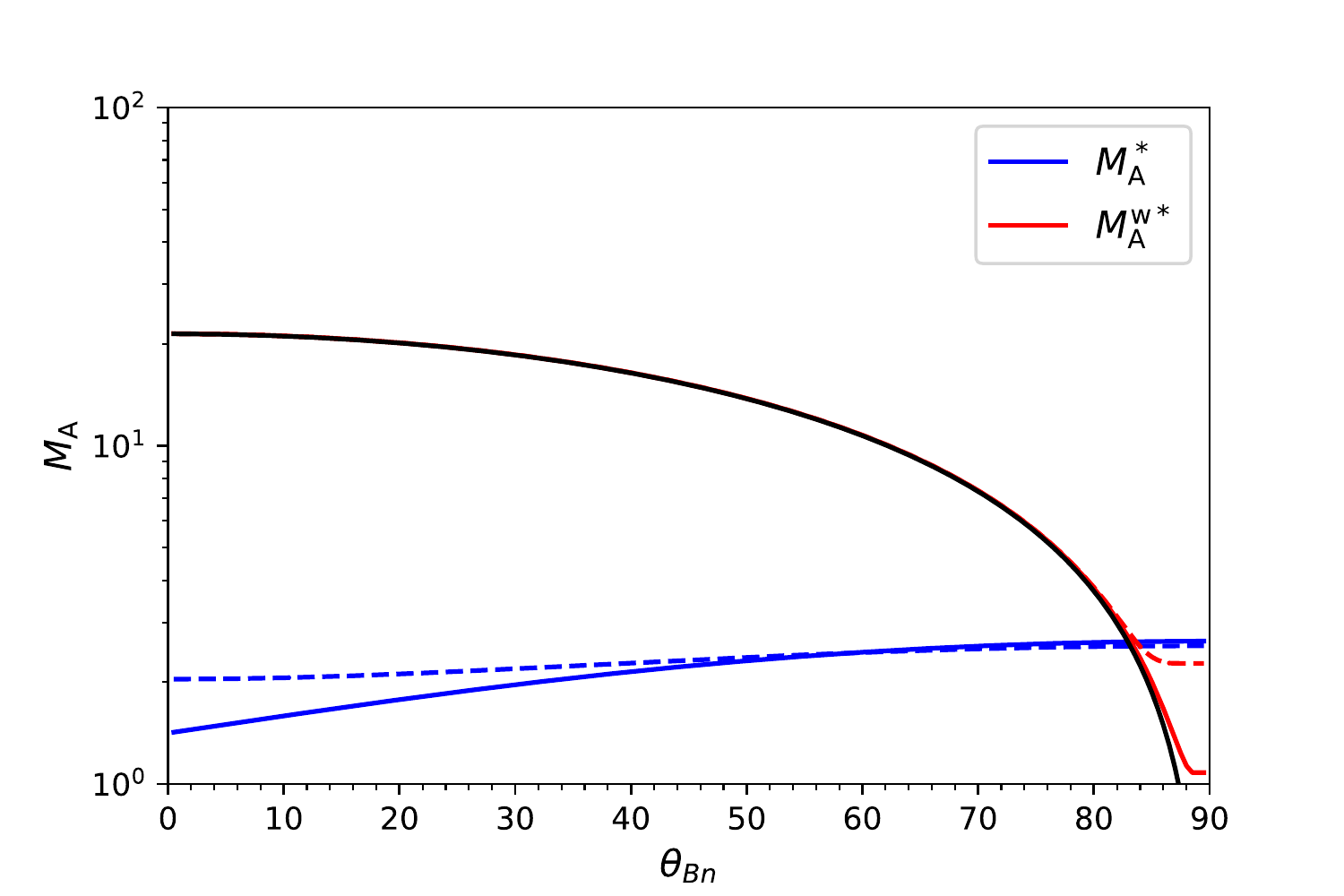}
    \caption{Critical \Alfven Mach numbers as functions of $\thetabn$. The blue and red curves show the critical Mach number $\MAc$ and $\MAwc$, respectively. The solid and dashed lines are for $\beta = 0.2$ and $5$. The black solid line shows the approximate whistler critical Mach number $\MAc \approx (m_i/m_e)^{1/2} \cos \thetabn/2$.}
    \label{fig:critical_ma}
\end{figure}

In short, in the absence of viscosity, resistivity and thermal conductivity can sustain only a weak shock. The evolutionary condition for a fast-mode MHD shock indicates that a finite viscosity is needed for a stable shock structure when the downstream sound speed is greater than the downstream flow speed. The condition defines the so-called critical Mach number, which may be calculated using the MHD Rankine-Hugoniot relations. In terms of the fast-magnetosonic Mach number, it is at most $\sim 2.7$ for a purely perpendicular shock in the low-beta limit and mostly below $\sim 2$ for finite beta plasmas $\beta \sim 1$ \citep{Edmiston1984,Kennel1985}. As we will discuss in the next subsection, the super-critical shock is characterized by the presence of the reflected ions that provide an effective viscosity. Since we will use the \Alfven Mach number $\MA$ and the obliquity $\thetabn$ throughout this paper instead of the fast-magnetosonic Mach number, it is instructive to introduce the corresponding critical \Alfven Mach number $\MAc$. \figref{fig:critical_ma} shows $\MAc$ as a function of $\thetabn$ for two plasma betas $\beta = 0.2, 5$ in solid and dashed lines, respectively. As one might expect, the critical \Alfven and fast-magnetosonic Mach numbers are more or less similar unless $\beta \gg 1$. We thus see that the majority of shocks seen in heliospheric and astrophysical applications are in the supercritical regime. Possible exceptions are weak interplanetary shocks or planetary bow shocks in unusual solar wind conditions \citep{Mellott1984a} and galaxy merger shocks in high-$\beta$ intracluster media ($\beta \gg 1$) \citep{Ha2018}.

Another important aspect that can be studied with the fluid description is the dispersive effect arising from the decoupling between the ion and electron dynamics. A finite-amplitude compressional wave will naturally steepen to convert its energy to short-wavelength fluctuations. If the dissipation is sufficiently weak, the steepening will proceed until generating modes with wavelengths comparable to the ion inertial length. At this scale, there appear finite deviations of the phase and group velocities from that of fast-mode in MHD. Therefore, dispersive wave trains will be emitted from the shock. Depending on the wave propagation speed, either leading (i.e., emitted toward upstream) or trailing (i.e., emitted toward downstream) wave trains may appear in the shock profile. It is then natural to introduce another critical Mach number that is associated with the dispersive nature of the fast-mode wave. The so-called whistler critical Mach number denoted by $\MAwc$ is defined as the threshold beyond which dispersive whistler precursor waves cannot propagate upstream \citep{Kennel1985}. In other words, it is determined by the fastest wave phase speed on the linear whistler-mode branch, which is approximately given by $\omega/k \VA \approx \sqrt{m_i/m_e} \cos \thetabn / 2 $ at around $kc/\wpe \approx 1$. The approximate expression indicates that the whistler critical Mach number is given by $\MAwc \approx (m_i/m_e)^{1/2} \cos \thetabn / 2$. \figref{fig:critical_ma} shows the exact definition of $\MAwc$ as a function of $\thetabn$ for two plasma betas $\beta = 0.2, 5$ obtained numerically with the red solid and dashed lines, respectively. It is clear that the approximate expression shown in the solid black line matches very well with the numerical results except for $\thetabn \approx 90\degr$. The rather strong dependence on $\thetabn$ can make Earth's bow shock (with $\MA \sim$ 5--10) both sub-critical or super-critical with respect to $\MAwc$, depending on $\thetabn$. Statistical analysis of Earth's bow shock confirmed that short-wavelength whistler precursors are seen only in the immediate upstream region of whistler sub-critical shocks $\MA \lesssim \MAwc$, which is consistent with the definition \citep{Oka2006}. On the other hand, an unexpected finding by the same authors was that efficient nonthermal electron acceleration had been observed preferentially at whistler-super-critical shocks. We should remind the readers that the definition of the whistler critical Mach number has formally nothing to do with particle acceleration efficiency. In Section \ref{sec:3}, we shall discuss the importance of the parameter $\MAHTF = \MA/\cos\thetabn$ in regulating the electron acceleration efficiency.

\subsection{Ion Reflection} \label{sec:2-ion-reflection}
The fluid theory implies the need for an effective viscosity for the super-critical collisionless shocks but does not predict how it is actually realized. Early in-situ spacecraft observations of Earth's bow shock, as well as kinetic numerical simulations, found that the heating at the shock occurs in association with the ion reflection \citep{Paschmann1981,Paschmann1982,Sckopke1983,Leroy1981,Leroy1982}. Namely, a fraction of ions flowing into the shock is reflected back upstream by the electromagnetic structure of the shock front. At a perpendicular shock, the reflected ions gain a significant amount of energy through traveling along the upstream motional electric field ($\bm{E} = - \bm{U}/c \times \bm{B}$) during their partial gyration in the immediate upstream. It is easy to confirm that the typical energy gain during the gyration is comparable to the upstream ion flow kinetic energy. Consequently, the reflected ions, when transmitted through the downstream region, can provide the thermal pressure that is sufficient to satisfy the pressure balance across the shock. As the obliquity $\thetabn$ decreases, the reflected ions tend to back-stream along the upstream magnetic field for a distance much longer than the gyroradius. Such a field-aligned back-streaming ion beam is generated typically at quasi-parallel shocks $\thetabn \lesssim 45 \degr$. The ion beam is expected to excite low-frequency \Alfvenic fluctuations by interacting with the upstream plasma via ion cyclotron resonance. Therefore, the quasi-parallel shock is accompanied by large-amplitude MHD turbulence in the pre-shock medium over a substantial distance, whereas the quasi-perpendicular shock has a relatively thin transition layer with a quiet upstream. The back-streaming ions ahead of a quasi-parallel shock will be scattered by the waves and isotropized in the local plasma rest frame. They are then eventually transported back to the downstream region, where they constitute the major portion of the downstream thermal pressure. Therefore, the reflection and subsequent energization of incoming ions at the shock is the essential ingredient of collisionless shock heating and may be considered as the means to provide an effective viscosity.

\begin{figure}[tbp]
    \centering
    \includegraphics[width=1.00\textwidth]{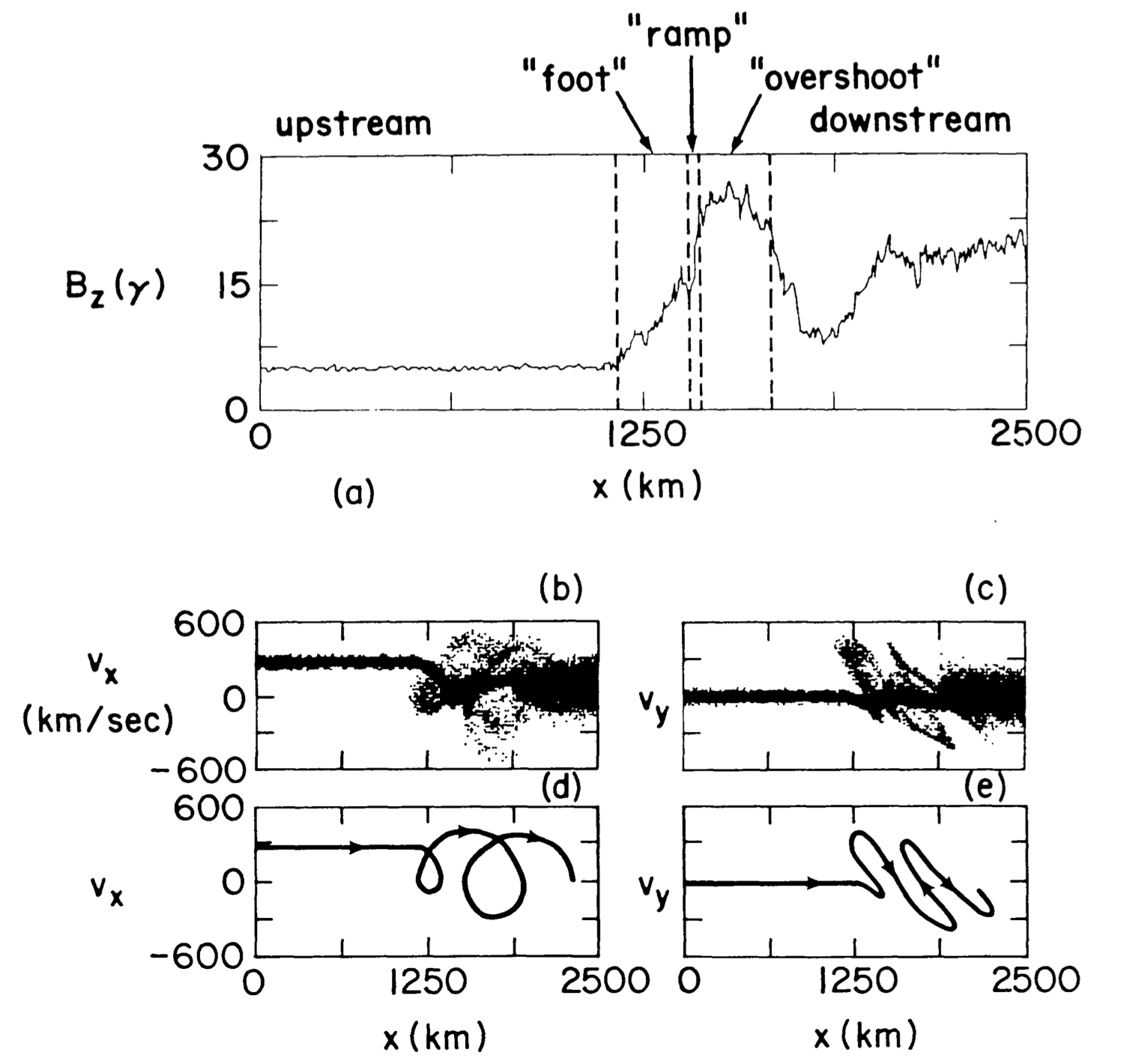}
    \caption{Kinetic hybrid (particle ions and fluid electrons) simulation for a perpendicular shock: (a) magnetic field spatial profile, (b) phase space particle distribution in $x - v_x$, (c) $x - v_y$, (d) typical reflected ion trajectory in $x - v_x$, (e) $x - v_y$. This figure is reproduced from \citet{Wu1984a}.}
    \label{fig:wu1984}
\end{figure}

Let us now consider a simplified scenario of how the ion reflection is realized at a perpendicular shock. The ion reflection at the shock front takes place as a combined effect of the magnetic deflection and the electrostatic potential deceleration. The compressed magnetic field at the shock tends to decelerate the upstream particles streaming into the shock. While electrons are more easily decelerated by the magnetic compression, ions will penetrate deeper into it due to their larger inertia. An electrostatic potential will then develop in the shock structure as a result of charge separation, which is called the cross-shock electrostatic potential. Using the generalized Ohm's law and dropping the finite electron inertia effect, the electrostatic potential in NIF can be formally written as follows \citep{Leroy1981,Leroy1982}
\begin{align}
    \phinif (x) = \int_{-\infty}^{x}
    \left[
        \frac{V_{i,y}}{c} B_z +
        \frac{1}{n e} \frac{\partial}{\partial x}
        \left( \frac{B_z^2}{8\pi} + p_e \right)
    \right] dx',
    \label{eq:phinif}
\end{align}
where the integration is taken along the shock normal (i.e., $x$) direction, and the magnetic field is assumed to be in the $z$ direction. If the ion bulk velocity in the $y$ direction $V_{i,y}$ is ignored, the potential develops due to the magnetic and electron thermal pressure gradient. As the shock compression proceeds, the pressure gradient increases, and at some point, a fraction of ions (in the tail of the upstream thermal distribution with relatively small kinetic energies in the shock frame) will be reflected back by the potential. The reflected ions will be accelerated in the $y$ direction in the upstream and contribute to the development of the potential through the first term. As a result, a slightly compressed region in front of the main shock ramp forms and is called the foot region. It is clear that the size of the foot region is characterized by the reflected ion gyroradius $\sim \Us/\wci$. The gyromotion of the reflected ions under the convection by the $\bm{E} \times \bm{B}$ flow produces the characteristic overshoot-undershoot magnetic structure with the scale size again given by the gyroradius. \figref{fig:wu1984} reproduced from \citet{Wu1984a} shows the super-critical shock structure consisting of foot-ramp-overshoot-undershoot reproduced by a hybrid simulation (with kinetic ions and a massless charge-neutralizing electron fluid). The characteristic magnetic-field profile is clearly associated with the typical reflected ion trajectory shown in panels (d) and (e). The magnetic structure and the ion distribution function measured by in-situ spacecraft at Earth's bow shock are fully consistent with the kinetic modeling.

While the magnitude of the cross-shock potential cannot be estimated without knowing the concrete shock internal profile, it will be comparable to the upstream ion flow kinetic energy $\Einif = m_i U_s^2/2$ so that the shock can reflect a non-negligible fraction of the upstream ions. Indeed, kinetic simulation results indicate $\phinif / \Einif \approx 0.1\mathrm{-}1$ \citep{Leroy1982,Thomsen1987a}. Although it is well established that the cross-shock potential plays an important role in the collisionless shock dynamics, we should always keep in mind that the electric field is a frame-dependent quantity. For instance, one of the important electron heating mechanisms at shocks is the energization associated with the cross-shock potential. However, the important quantity may not necessarily be the one measured in NIF. We will discuss this issue in more detail in the next subsection.

\subsection{Electron Heating} \label{sec:2-electron-heating}

We see from the discussion in the previous subsection that the scale size of the super-critical shock structure given by the ion gyroradius will be much larger than the typical gyroradii of thermal electrons. One may thus anticipate that thermal electrons will behave adiabatically in such a laminar shock structure in the sense that the first adiabatic invariants are approximately conserved. Indeed, electron heating at super-critical Earth's bow shock is often weak compared to ion heating.


Let us now first discuss the electron dynamics under the adiabatic approximation. Conservation of the first adiabatic invariant for individual electrons indicates that $T_{e,\perp}/B \approx \textrm{const.}$ is satisfied even within the shock structure, where $T_{e,\perp}$ is the electron temperature perpendicular to the magnetic field. This implies that the perpendicular electron temperature increase will be limited by the magnetic compression ratio across the shock. At a high Mach number shock, this adiabatic heating is clearly insufficient to provide the temperature downstream of the shock as required by the Rankine-Hugoniot relationship ($T \propto \Us^2$), and the deficit must be supported by ions. On the other hand, in general, the cross-shock electrostatic potential may have a field-aligned component at an oblique shock, which can influence the electron dynamics in the parallel direction. Since the potential develops to reflect positively charged ions, it tends to accelerate the thermal electrons into the downstream as a bulk. Therefore, the cross-shock potential may be an important source of electron energization at collisionless shocks.


It is, however, important to understand that the amount of energy that is transferred to the electron through the parallel potential drop is not directly determined by $\phinif$ because of its frame dependence. Although it is possible to calculate the electron energy change through the shock in NIF, one has to take into account the motional electric field that tends to compensate for the energy gained by $\phinif$ \citep[e.g.,][]{Goodrich1984}. Instead, HTF has conventionally been utilized as it substantially simplifies the analysis of the charged particle interaction with the shock. The absence of the motional electric field in this frame indicates that the energy gain by the particles should be determined solely by the potential drop across the shock measured in HTF $\phihtf$. While the cross-shock potential can be the source of electron heating, the actual electron energization will involve irreversible heating associated with wave-particle interaction. In-situ observations of the electron velocity distribution function (VDF) within the shock transition layer clearly demonstrated that the upstream thermal electrons are accelerated as a bulk toward downstream, but at the same time, they tend to diffuse in velocity space as the spacecraft penetrates deeper into the shock \citep{Feldman1982a,Feldman1983a}. The downstream VDF eventually becomes a flat-topped distribution, which is nearly flat at low energy and connected smoothly to a power-law tail at high energy. The formation of the flat-top VDF has been considered as a result of microscopic electrostatic instabilities driven by the bulk electron beam (generated by the cross-shock potential) interacting with the ambient ions and electrons \citep{Scudder1986a,Thomsen1987b,Schwartz1988b}.


Unfortunately, however, the knowledge of $\phinif$ alone is insufficient to estimate $\phihtf$ and the resulting electron heating. The frame transformation of the normal component of the electric field from NIF to HTF involves the magnetic field component out of the coplanarity plane \citep{Thomsen1987a}. The non-MHD magnetic field component within the shock transition layer results from the dispersive nature of the fast-mode wave at the ion inertial scale. This makes it difficult to obtain an analytic estimate of the electron heating rate associated with the cross-shock potential unless the shock internal structure is appropriately reproduced with kinetic modeling, including the effect of the reflected ions. In-situ observations of the bow shock and relevant kinetic simulations found $\phihtf / \Einif \approx 0.1-0.3$ \citep{Thomsen1987b,Schwartz1988b}, which indicates a typical electron-to-ion temperature ratio of $T_e/T_i \sim 0.1$ in the downstream of the shock. However, the limited parameter range of the solar wind does not allow us to obtain how the temperature ratio possibly scales with the shock parameters.


The discussion so far assumes as if the collisionless shock were laminar at the electron scale, which is, however, not the case \citep[e.g.,][]{Wilson2010,Wilson2014,Goodrich2018,Hull2020}. In addition to the cross-shock potential, the strong perpendicular current associated with the steep magnetic gradient and the reflected ion beam are two major sources of microscopic plasma instabilities \citep{Wu1984a}. The instability condition and resulting heating efficiency will, in general, depend strongly on the shock parameters. It is important to note that the dynamics of collisionless shocks depend not only on macroscopic parameters (such as $\MA$, $\beta$, $\thetabn$) but also the ion-to-electron mass ratio $m_i/m_e$ as well as the electron plasma-to-cyclotron frequency ratio $\wpe/\wce$. In other words, the electron heating efficiency may differ substantially even with the same macroscopic shock parameters, which makes it difficult to obtain scaling laws of various heating processes from kinetic simulation results.

It might be more straightforward to obtain an empirical scaling law from observations of shocks under various conditions. Astrophysical observations of Balmer-line emissions from shocks in partially ionized media suggest that the temperature ratio may decrease as increasing the shock speed \citep[e.g.,][]{Ghavamian2013a}. Similar results have also been reported based on Cassini spacecraft measurements of Saturn's bow shock \citep{Masters2011a}. However, the physical mechanism of such dependence has not been understood very well. It might be possible that a significant fraction of energy is transferred to the nonthermal electron population rather than the thermal electrons at high Mach number shocks, which apparently reduces the electron temperature. Theory and kinetic simulation results presented in this review support this interpretation.


To summarize, it appears common that the electron heating at collisionless super-critical shocks is weaker than the ions, at least at Earth's bow shock under the average solar wind condition. The energization by the cross-shock potential in the parallel direction and the subsequent irreversible velocity space diffusion is a reasonable way to understand electron heating. There are certainly other possible heating mechanisms associated with microscopic plasma instabilities, which may or may not be activated depending on the shock parameters. In Section \ref{sec:4} and \ref{sec:5}, we will discuss the Buneman and Weibel instabilities, both of which require very high Mach numbers to be excited. The electron heating and nonthermal electron acceleration associated with these instabilities are important subjects of this paper.

\subsection{Electron Injection Problem} \label{sec:2-electron-injection}
The standard DSA theory for the acceleration of charged particles at shocks assumes that the accelerated particles diffusively bounce back and forth across the infinitesimally thin shock transition layer many times. Therefore, the accelerated particles must satisfy the following two conditions: (1) they should freely go across the shock front in both directions, (2) they should experience sufficiently strong pitch-angle scattering so that the momentum distribution becomes approximately isotropic in the fluid rest frame. In the following, we discuss these conditions for low-energy electrons.

Let us discuss the first condition (1). For this, it is crucial to understand the meaning of the assumption made by the standard DSA that the shock is infinitesimally thin. This implies that the finite thickness of the shock transition layer has to be smaller than a typical length scale of accelerated particles. One of the natural choices for the length scale is the gyroradius. If we assume that the shock thickness is given by the ion gyroradius determined by the upstream flow velocity, we immediately find that the minimum electron momentum required to satisfy the condition is given by
\begin{align}
    \frac{p}{m_e \VA} \approx \left( \frac{m_i}{m_e} \right) \MA.
    \label{eq:pinj_gyro}
\end{align}
This momentum has often been quoted as the injection threshold. The corresponding energy is typically $\sim 10$ MeV at SNR shocks and is orders of magnitude higher than the electron thermal energy in the downstream. \citet{Ellison1991}, for instance, argued that the acceleration of electrons can be treated exactly the same way as protons beyond this momentum. It does not mean, however, that this provides the strict threshold for DSA.

Indeed, one may adopt a somewhat different viewpoint. As we understand that the adiabatic theory provides a reasonable first-order approximation of the low-energy electron interaction with a collisionless shock, we can assume that they traverse an oblique (subluminal) shock rather freely from both sides. More specifically, an upstream electron has to satisfy a certain condition in pitch angle to go across the shock because otherwise, they will be reflected back by the magnetic mirror force at the shock. Similarly, electrons in the downstream need to have sufficiently high energies to escape away toward the upstream along the magnetic field line because of the presence of the cross-shock potential in HTF $\phihtf$. Since we will discuss the adiabatic theory in more detail in Section \ref{sec:SDA}, here, it suffices to say that the typical energy required is only a few times the potential energy $\phihtf$, which is roughly the electron thermal energy in the downstream. We see that the required energy with this argument is much smaller than that suggested by \eqref{eq:pinj_gyro}. As long as particles suffer pitch-angle scattering both in the upstream and downstream, DSA will be effective for low-energy electrons that traverse the shock in a fully adiabatic manner. Note that the adiabatic mirror reflection effect has intrinsically been integrated into DSA at an oblique shock \citep{Drury1983}. If the scattering is strong enough to validate the diffusion approximation for the spatial particle transport, the balance between diffusion and convection by the background plasma flow defines a characteristic length scale called the diffusion length. In Section \ref{sec:3}, we will show that it is actually the diffusion length rather than the gyroradius that has to be compared with the shock thickness.

The second condition (2) is actually closely related to the above discussion. It is reasonable to assume that a certain level of turbulence exists in the upstream and downstream regions of the shock. The turbulence may be pre-existing in the far upstream region or can be generated in the immediate upstream by the shock-accelerated particles. In both cases, the fluctuations will be transmitted through the shock, and the downstream plasma will also be turbulent. The question is whether the turbulence is capable of scattering low-energy electrons. In the classical understanding, the particles can be scattered efficiently through the cyclotron resonance. In the low-frequency limit, the resonance condition is written quite reasonably by $k r_g \sim 1$, where $r_g$ is the particle gyroradius. Therefore, the scattering of low-energy electrons requires small-scale fluctuations with a wavelength comparable to the electron gyroradius. It is well below the dissipation length scale of MHD turbulence, which is typically determined by the ion scale. This argument holds even if we consider the exact resonance condition $\omega = k v_{\parallel} + \wce$ with a right-hand polarized wave ($\omega > 0$) \citep{Amano2010}. Unless there exists a continuous energy injection, the small-scale fluctuations are subject to strong collisionless damping and will not be able to provide sufficient scattering for electron acceleration.

Therefore, we may consider that the minimum momentum for electron injection corresponds to the one at which the electron gyroradius becomes comparable to the dissipation length scale of turbulence. We define the wavenumber $k_{*}$ at which the turbulence spectrum will steepen because of the dissipation. It would be reasonable to consider that the thermal ion cyclotron damping roughly determines $k_{*}$. In a low-beta plasma, the dissipation scale is given by the ion inertial length $k_{*} c/\wpi \sim 1$, whereas the same argument leads to $k_{*} c/\wpi \sim \betai^{-1/2}$ in a high-beta plasma. Note that the latter condition indicates that the dissipation is controlled by the thermal ion gyroradius. The cyclotron resonance condition in the low-frequency limit $k_{*} r_g \sim 1$ indicates that the injection threshold momentum $p_\textrm{inj}$ is given by
\begin{align}
    \frac{p_\textrm{inj}}{m_e \VA} \approx
    \left( \frac{m_i}{m_e} \right)
    \left( \frac{k_{*} c}{\wpi} \right)^{-1},
    \label{eq:pinj_resonance}
\end{align}
with the dissipation scale determined by $k_{*} c/\wpi \sim \textrm{min}(1, \betai^{-1/2})$ \citep{Amano2022a}. It is seen that the estimate by \eqref{eq:pinj_resonance} can be much smaller than \eqref{eq:pinj_gyro} at high-$\MA$ shocks. For typical interstellar and interplanetary media, mildly relativistic energy, say $\sim 0.1-1$ MeV, is needed as the threshold energy for injection. It is interesting to note that, if we consider a parallel shock in which upstream MHD waves are produced by shock-reflected ions via the cyclotron resonance, the dominant wavenumber will be $k c/\wpi \sim 1/\MA$. Substituting this estimate to $k_{*}$, we recover \eqref{eq:pinj_gyro}. We may thus consider that the threshold momentum \eqref{eq:pinj_gyro} provides the most pessimistic scenario as it excludes any nonlinear cascade of the wave power spectrum.

As we have seen above, the first condition (1) is not necessarily an issue if the adiabatic approximation is appropriate for the low-energy electron interaction with the shock. On the other hand, the second one (2) places a much more stringent restriction on electron energy. There are potentially two different scenarios to resolve the electron injection problem. The first is to invoke a completely different particle acceleration mechanism other than DSA that can provide a sufficiently energetic seed population. The electron energization within the shock transition layer through the interaction with various microscopic plasma instabilities is a probable candidate. The second option is to persevere with DSA but consider a generation mechanism of high-frequency waves that can resonantly scatter low-energy electrons. In other words, if the dissipation length scale is substantially smaller than the above estimate (i.e., $k_{*} c/\wpi \gg 1$), DSA will become effective for much lower energy electrons. The strong damping expected for high-frequency and short-wavelength waves indicates that there must be an instability that can overcome the damping. Self-generation mechanisms of relevant waves by the accelerated electrons themselves have been proposed as possible injection mechanisms \citep{Levinson1992a,Levinson1994,Levinson1996,Amano2010}. The novel injection mechanism via SSDA that will be described in Section \ref{sec:3} can be considered as a hybrid scenario between the two options \citep{Katou2019,Amano2020}.

Unless there exists an efficient injection mechanism, the efficiency of sub-relativistic electron acceleration at shocks is, in general, considered to be much lower in theory. Observations of shocks within the heliosphere are largely consistent with this theoretical expectation \citep{Lario2003,Dresing2016}. It is nevertheless important to mention that suprathermal electrons have sometimes been measured at and around Earth's bow shock. Although the energy range is limited to a few hundreds of keV at most \citep[e.g.,][]{Anderson1981,Gosling1989,Oka2006,Wilson2016}, it is indeed the key energy range for electron injection. We thus believe that understanding the acceleration of electrons at Earth's bow shock, in particular its shock-parameter dependence, will provide the clue to resolve the problem of electron injection at collisionless shocks.


\clearpage
\section{Stochastic Shock Drift Acceleration} \label{sec:3}

\subsection{Motivation}

This section discusses the particle acceleration mechanism called stochastic shock drift acceleration (SSDA), which has recently been recognized as a promising candidate for electron injection \citep{Katou2019,Amano2020,Amano2022a}. SSDA has been constructed on top of the classical shock drift acceleration (SDA) by adding the effect of stochastic pitch-angle scattering.

It is instructive to begin our discussion with the motivation to consider SDA as a building block for constructing a theory of electron injection. SDA was first proposed as a particle acceleration mechanism that may explain the observations of energetic electrons in the upstream of Earth's bow shock \citep{Wu1984b,Leroy1984a}. These particles are streaming away from the shock along the local magnetic field lines, with the most energetic population being observed when the local field line is nearly tangent to the curved bow shock \citep{Anderson1979,Anderson1981}. In other words, they are likely generated at a nearly perpendicular portion of the bow shock. Through extensive theoretical and experimental studies, the idea of SDA has been well established \citep[e.g.,][]{Krauss-Varban1989b,Krauss-Varban1991,Vandas2001}. While the theory is qualitatively consistent with observations, it was shown that the adiabatic SDA alone cannot quantitatively explain measured power-law spectra and large fluxes of high-energy electrons \citep{Vandas2001}.

As we see below, SSDA introduces a stochastic nature into the intrinsically deterministic SDA. This enhances the efficiency of particle acceleration, enabling the injection into diffusive shock acceleration (DSA) at high Mach numbers. We also discuss a unified model that combines SSDA and DSA under the diffusion approximation. It consistently describes both the injection at low energy and subsequent acceleration at high energy at the same time.

\subsection{Adiabatic Shock Drift Acceleration} \label{sec:SDA}

\begin{figure}[tbp]
    \centering
    \includegraphics[width=1.00\textwidth]{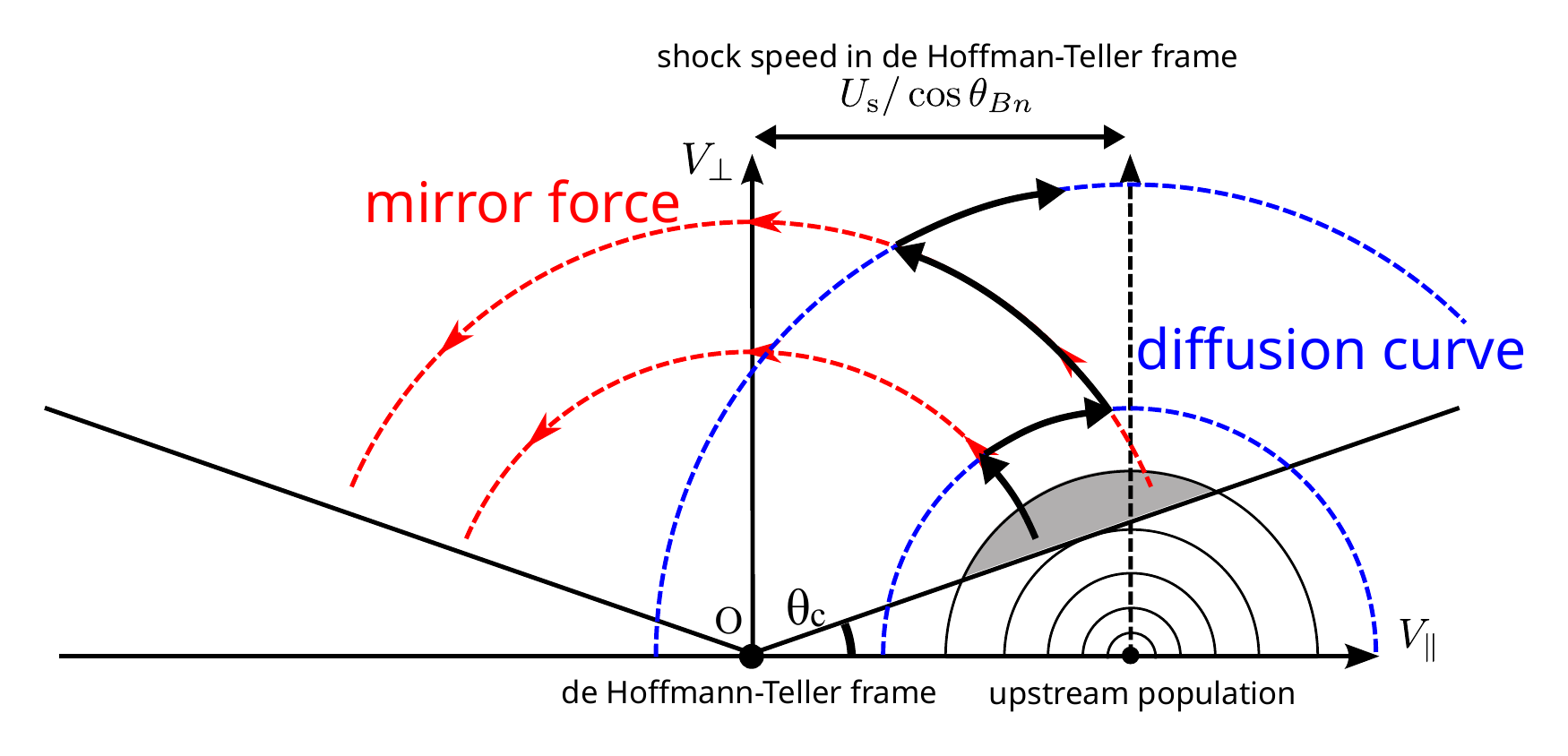}
    \caption{Schematic illustration of particle trajectory in velocity space expected by SSDA. The upstream thermal population is shown in thin black contours. In the absence of scattering, particles in the shaded area are subject to the adiabatic mirror reflection (SDA) and accelerated along the red dashed lines. Pitch-angle scattering induces diffusion of particles along the diffusion curves shown with the blue dashed lines. The thick black arrows represent the expected particle trajectory for SSDA.}
    \label{fig:ssda}
\end{figure}

We first briefly review the classical SDA under the adiabatic approximation. Readers interested in more details may refer to the original papers \citep{Wu1984b,Leroy1984a}. Let us consider the shock as a smooth magnetic field compression over the spatial scale comparable to the reflected ion gyroradius. A small-gyroradius low-energy electron in the upstream flowing into the shock transition layer will interact with the compressed magnetic field structure with keeping the adiabatic moment approximately constant. We first consider the dynamics of electrons in the de Hoffmann-Teller frame (HTF). In this particular frame of reference, the electric field perpendicular to the magnetic field vanishes. If we further ignore the cross-shock electrostatic potential (i.e., the parallel electric field), the particle energy in HTF will be conserved. Therefore, the particle trajectory is characterized only by the pitch angle $\alpha$ defined also in HTF. A low-energy incoming electron will be reflected back upstream via the magnetic mirror force if the pitch angle is greater than the loss-cone angle $\alpha \geq \theta_{\rm c} = \sin^{-1} \left( \sqrt{B_0/\Bmax} \right)$, where $\Bmax$ denotes the maximum magnetic field strength within the shock profile. The resulting momentum gain measured in the upstream plasma frame is, on average, given by $\Delta p \sim 2 m_e \Us/\cos \thetabn = 2 m_e \VA \MAHTF$. Therefore, a substantial energy gain is expected at a nearly perpendicular shock with a very large HTF \Alfven Mach number $\MAHTF$. The particle trajectory in velocity space expected for SDA is schematically illustrated with red curves in \figref{fig:ssda}.

The mirror reflection of an electron with a typical velocity $v$ in the upstream rest frame occurs only when $v/(\Us/\cos \thetabn) \gtrsim \sin \theta_{\rm c}$, which may be rewritten as $v \gtrsim \Us / (2 \cos \thetabn)$ (for $\Bmax/B_0 \sim 4$). Therefore, the reflection will not take place if the initial electron energy in the upstream rest frame is not large enough. For a particle of given energy, the reflection condition becomes more and more difficult to satisfy as increasing $\MAHTF$ where the energy gain is more efficient. While a nearly perpendicular shock with $\thetabn$ close $90 \degr$ is favorable for higher energy gain, the number of particles that satisfy the reflection condition will rapidly decrease if the upstream electron distribution is described by a thermal Maxwellian. For this reason, earlier works considered the reflection of pre-existing suprathermal electrons at a nearly perpendicular shock to explain the highly energetic electrons observed in the upstream of the bow shock. The sensitivity on $\thetabn$ can be relaxed by adding the effect of pitch-angle scatterings in SSDA.


A finite HTF cross-shock potential does not affect the momentum gain for the reflected particles but does change the reflection condition. Since the potential accelerates the electrons toward downstream, it tends to increase the threshold pitch angle at low energies where the kinetic energy in HTF is comparable to the potential energy. On the other hand, its effect will not be significant once the electrons are accelerated beyond a few times the potential energy.


While HTF is a convenient frame for theoretical analysis, it is also important to understand the particle energization mechanism as seen in a different reference frame because neither observations nor simulations are usually in HTF. \citet{Krauss-Varban1989a} showed that the energy gain mechanism seen in NIF can be understood by the gradient-B drift in the direction (or the anti-parallel direction for a negative charge) of the convection electric field $\bm{E} = - \bm{U} \times \bm{B} / c$, which is constant for a stationary MHD shock. Therefore, the particle energy gain is proportional to the travel distance along the electric field. In other words, the electron energy gain per unit time during the interaction with the shock is given by
\begin{align}
    \frac{d \epsilon}{d t} = -e \bm{v}_{\nabla B} \cdot \bm{E}
    = \frac{m v_{\perp}^2}{2 B} c \left( \bm{b} \times \nabla \ln B \right) \cdot \bm{E}
\end{align}
where $\bm{v}_{\nabla B}$ is the gradient-B drift velocity. If we assume that the magnetic-field gradient is more or less constant within the shock transition layer (i.e., $d \ln B / dx \approx \textrm{const}$), the acceleration timescale $\tau_\mathrm{acc, SDA}$ may be estimated by
\begin{align}
    \frac{1}{\tau_{\mathrm{acc, SDA}}} = \frac{1}{\epsilon} \frac{d \epsilon}{d t}
    \sim \Us \frac{d \ln B}{d x}
    \sim \wci
\end{align}
where we have assumed that the thickness of the shock transition layer $\Ls$ is given by $\Ls \sim (d \ln B/dx)^{-1} \sim \Us/\wci$ and $v \sim v_{\perp}$. We note that the timescale of the upstream plasma convected through the shock transition layer is roughly given by $\sim \wci^{-1}$. Therefore, particles transmitted through the shock gain energy comparable to the initial energy, which is consistent with the adiabatic perpendicular energy gain to conserve the first adiabatic invariant. On the other hand, the reflected particles gain more energy because they stay longer within the shock transition layer. SSDA will further increase the residence time by introducing pitch-angle scattering, which diffusively confines the accelerated particles within the particle acceleration region.

\subsection{Effect of Scattering}

In-situ observations of Earth's bow shock by NASA's MMS spacecraft clearly identified the cyclotron resonant interaction between low-energy electrons and high-frequency whistler-mode waves \citep{Oka2017}. Electron scattering associated with low-frequency whistler waves has also been discussed \citep{Riquelme2011,Oka2019}. Since the pitch-angle scattering by the wave-particle interaction violates the assumption of adiabatic approximation made in SDA, the theory needs revision.

\citet{Katou2019} proposed SSDA as a model that takes into account the effect of pitch-angle scattering into the adiabatic SDA. \figref{fig:ssda} illustrates the mechanism of particle acceleration by SSDA. If the wave phase speed is negligible compared to the individual particle velocity (which is a good approximation for suprathermal electrons interacting with whistler waves at the bow shock), one may reasonably assume that the scattering conserves the particle energy in the plasma rest frame. In other words, the particle motion in velocity space interacting with the wave is constrained on the iso-energy contour, which is called the diffusion curve shown in blue in \figref{fig:ssda}. Recall that the timescale of SDA is on the order of $\wci^{-1}$, which will be much longer than the pitch-angle scattering timescale. Therefore, the pitch-angle scattering can be recognized as a random perturbation along the diffusion curve on the zeroth-order trajectory in velocity space predicted by the adiabatic SDA (shown in red). We see that particles scattered toward downstream will finally gain more energy than in the absence of the scattering (both in the plasma rest frame and HTF). In this case, the scattering prohibits the particle escape toward upstream and increases the particle residence time within the shock transition layer. Since the particle experiences a nearly constant rate of energy gain in the acceleration region, the increased residence time will naturally result in an increased energy gain.

It is important to point out that SSDA is a local process in the sense that the particle acceleration completes within the shock transition layer. This makes SSDA advantageous as an electron injection mechanism for the following two reasons: (1) The acceleration time is much faster than DSA because there is no need for particles to travel long distances both in the upstream and downstream as required by DSA. (2) Intense high-frequency wave activity is needed only within the shock transition layer and its close vicinity, where the free energy for the wave excitation is most abundant.


\subsection{Observational Evidence}

The question is obviously whether the strong scattering assumption made in the theory is realistic or not. Using MMS observations at Earth's bow shock, \citet{Amano2020} confirmed the weak pitch-angle anisotropy and the intense high-frequency wave activity, both of which are qualitatively consistent with the theory. Note that, in the immediate upstream solar wind, the high-frequency wave intensity was much lower. Accordingly, the pitch-angle anisotropy was larger. This implies that the energetic electrons were produced locally within the shock transition layer. However, a more quantitative test requires further theoretical development.

If the scattering is strong enough such that the pitch-angle distribution becomes nearly isotropic, the isotropic part of the electron distribution function $f(x,p)$ evolves according to the following diffusion-convection equation in HTF \citep{Amano2022a}
\begin{align}
	\frac{\partial f}{\partial t} +
	V_{x} \frac{\partial f}{\partial x} +
	\frac{1}{3} V_{x}
	\left(
		\frac{\partial \ln B}{\partial x} - \frac{\partial \ln V}{\partial x}
	\right)
	\frac{\partial f}{\partial \ln p} =
	\frac{\partial}{\partial x}
	\left(
		\kappa_{xx} \frac{\partial f}{\partial x}
	\right),
	\label{eq:diffusion-transport}
\end{align}
with which we can model the electron acceleration within the shock transition layer. Note that $x$ is the shock normal direction, $p$ is the momentum in the plasma rest frame, $V = V(x)$ is the field-aligned plasma flow velocity, $B = B(x)$ is the magnetic field strength, $\theta = \theta(x)$ is the magnetic field orientation, $\kappa = \kappa(x)$ is the parallel diffusion coefficient. In addition, we use the notation $V_{x} = V \cos \theta$ and $\kappa_{xx} = \kappa \cos^2 \theta$.

If the particle transport indeed obeys Eq.~(\ref{eq:diffusion-transport}), the steady-state solution with appropriate boundary conditions indicates that the particle flux at the shock will increase nearly exponentially over the spatial scale $\ldiff = \kappa_{xx}/V_{x} = \kappa \cos^2 \thetabn / \Us$, which is called the diffusion length. The diffusive particle confinement considered in SSDA requires the condition $\ldiff / \Ls \lesssim 1$. Such exponential flux profiles consistent with the theory have been seen in observations \citep[e.g.,][]{Oka2017}. \citet{Amano2020} performed a more quantitative test using the energy dependence of $\kappa$, which was determined by fitting the particle intensity profiles with an exponential function. The condition for SSDA $\ldiff / \Ls \lesssim 1$ may be rewritten in terms of particle energy $E$ as follows
\begin{align}
    \frac{D_{\mu\mu}}{\wce} \gtrsim
    \frac{1}{6 \eta}
    \left( \frac{m_e}{m_i} \right) \left( \frac{E}{\Eehtf} \right),
    \label{eq:threshold}
\end{align}
where $\Eehtf = m_e (\Us/\cos \thetabn)^2 / 2$ corresponds to the electron kinetic energy defined with the HTF shock speed and $\eta = \Ls/(\Us/\wci)$ is the normalized thickness of the shock transition layer. The pitch-angle scattering rate $D_{\mu\mu} = D_{\mu\mu}(E)$ is related to the diffusion coefficient via $\kappa = v^2/(6 D_{\mu\mu})$. The theory indicates that the particle acceleration by SSDA will take place only in the energy range where Eq.(\ref{eq:threshold}) is satisfied. Otherwise, no particle acceleration is expected, indicating the formation of the cutoff in the high-energy part of the energy spectrum.

\begin{figure}[tbp]
    \centering
    \includegraphics[width=1.00\textwidth]{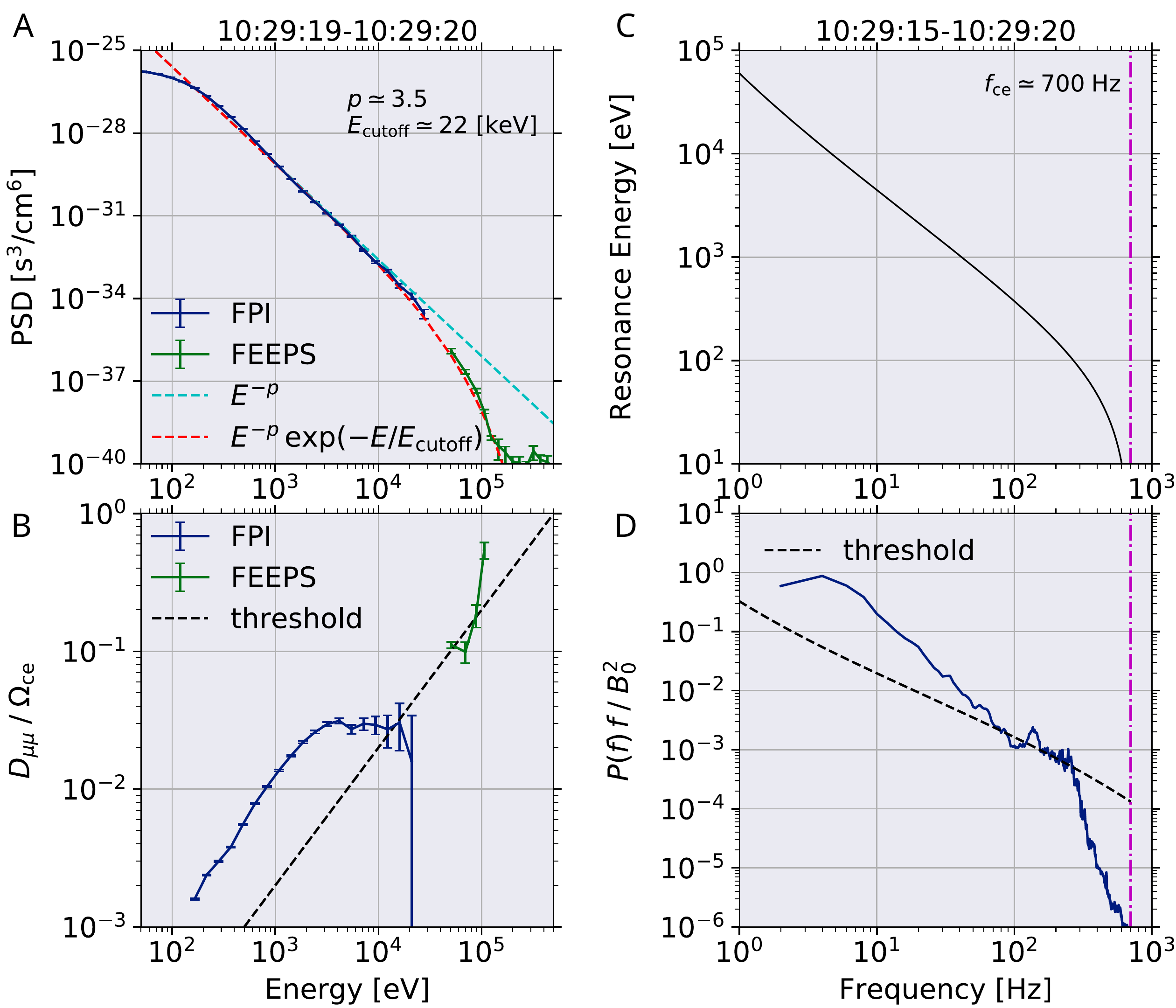}
    \caption{Comparison between theory and in-situ observations: (A) Electron phase space density as a function of energy, (B) scattering rate estimated from particle measurements, (C) relation between the wave frequency and the cyclotron resonant electron energy, (D) magnetic field fluctuation power spectrum. The black dashed lines in (B) and (D) represent the theoretical threshold corresponding to the right-hand side of \eqref{eq:threshold}. This figure is reproduced from \citet{Amano2020}.}
    \label{fig:amano2020a}
\end{figure}

\figref{fig:amano2020a} shows the summary of the comparison between the theory and observations by MMS \citep{Amano2020}. The electron energy spectrum shown in the panel (A) has a clear power-law part followed by a high-energy cutoff approximately at $\sim 20$ keV. The panel (B) shows the energy dependence of $D_{\mu\mu}$. The theoretical threshold given by the right-hand side of \eqref{eq:threshold} is shown with the dashed black line. Comparison between the panels (A) and (B) confirms that the power-law part corresponds to the energy range where \eqref{eq:threshold} is satisfied. The high-energy cutoff in the spectrum at $\sim 20$ keV corresponds to the energy where the scattering rate becomes equal to the theoretical threshold \eqref{eq:threshold}. It is important to note that the high-energy cutoff in theory is uniquely determined only by the single parameter $D_{\mu\mu}$ for the given shock speed $\Us$ and the obliquity $\thetabn$. Therefore, the agreement between the predicted and observed cutoff energy should be recognized as solid evidence for the theory.

Essentially the same argument can be tested with the wave measurement as well because the threshold scattering rate may be converted to the one in the wave intensity using the quasi-linear theory \citep{Katou2019,Amano2020}. The panel (C) shows the relation between the wave frequency and the cyclotron resonance energy, while (D) displays the measured wave intensity with the theoretical threshold also shown in the dashed black line. One can confirm that the measured wave intensity is comparable to or larger than that required by theory up to $\sim 300$ Hz. \citet{Amano2020} confirmed that the high-frequency waves ($\gtrsim 100$ Hz) correspond to quasi-parallel propagating whistler-mode waves that can scatter $0.1\mathrm{-}1$ keV electrons. This indicates that SSDA can successfully operate for suprathermal electrons with energies above $\sim 0.1$ keV where the power-law starts to develop. Note that at lower frequencies or correspondingly at higher energies, the wave intensity is always much larger than the threshold. However, since the scattering rate will obviously have an upper limit below the most optimistic Bohm scattering rate $D_{\mu\mu} \sim \wce$ at $\delta B/B_0 \sim 1$, the quasi-linear theory will break down at some point. This implies that the scattering rate should saturate at high energy, as is seen in the panel (B) above a few keV.

\subsection{Signatures in Kinetic Simulations}


\citet{Matsumoto2017} statistically analyzed trajectories of accelerated electrons obtained by a fully 3D PIC simulation of an oblique high-Mach-number quasi-perpendicular shock with $\MA = 20.8, \thetabn = 74\degr, \betai = \betae = 0.5, \wpe/\wce = 10, m_i/m_e = 64$. They found that the accelerated electrons suffer shock surfing acceleration (SSA) first in the leading edge of the shock. The pre-accelerated electrons are then energized further by SSDA in the deeper shock transition layer. Here we concentrate on SSDA, as we present a detailed discussion on SSA later in Section \ref{sec:4}.

\begin{figure}[tbp]
    \centering
    \includegraphics[width=1.00\textwidth]{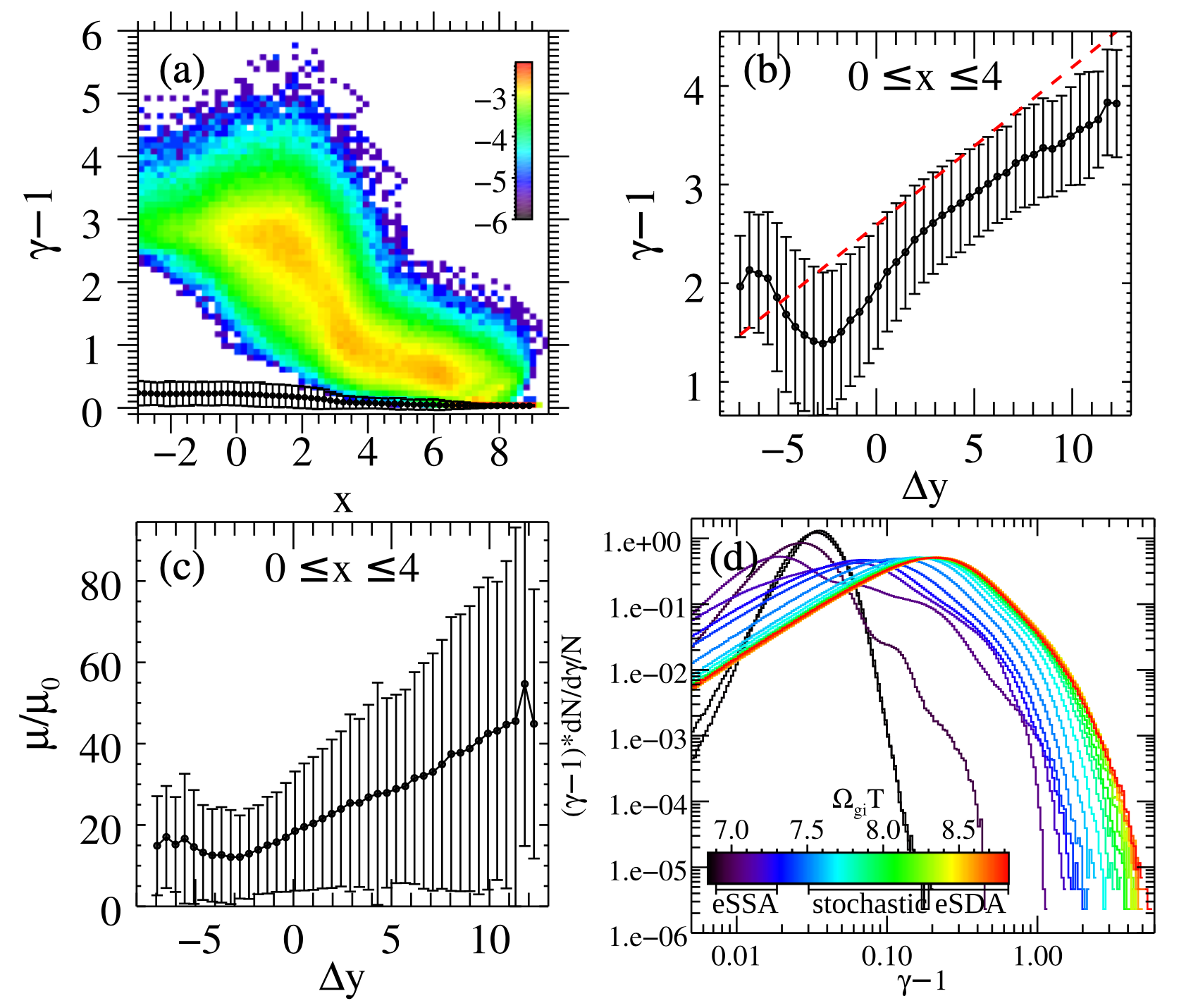}
    \caption{Statistical analysis of energetic electrons seen in the fully 3D simulation by \citet{Matsumoto2017}: (a) Probability distribution function of energetic electrons in phase space, (b) energy of energetic electrons as a function of displacement in $y$ direction, (c) the first adiabatic invariant of energetic electrons as a function of displacement in $y$ direction, (d) time evolution of the energy spectrum calculated from the pre-selected self-consistent tracer particles. The average behavior of the thermal electrons is also shown in the panel (a) in black. This figure is reproduced from \citet{Matsumoto2017}.}
    \label{fig:matsumoto2017a}
\end{figure}

\figref{fig:matsumoto2017a} shows the statistical analysis of the most energetic particle population seen in the simulation \citep{Matsumoto2017}. They found that the probability distribution of energetic particles in phase space, shown in color in the panel (a), is distinct from the thermal ones (shown in black). The panel (b) shows the energy gain as a function of the particle travel distance in the direction opposite to the convection electric field ($\Delta y$), in which a clear linear trend consistent with the SDA theory is seen. The panel (c) demonstrates that the accelerated particles exhibit a systematic increase (with a substantial variance) in the adiabatic invariant as a function of the travel distance. The result suggests that the energetic electrons suffer strong scattering. In other words, the particle acceleration is consistent with SSDA in that the energy gain itself is the same as SDA, and the accelerated particles are confined within the shock transition layer by strong scattering. The panel (d) displays the time evolution of the energy spectrum constructed from self-consistent tracer particles distributed initially at the same $x$ position in the upstream region of the shock. The early phase energization in a short time period is due to SSA, while the gradual formation of the power-law spectrum in the late time is considered as the result of SSDA. Note that the scattering, in this case, is not due to whistler waves but large-amplitude Weibel-generated turbulence, which will be discussed in more detail in Section \ref{sec:5}

\begin{figure}[tbp]
    \centering
    \includegraphics[width=1.00\textwidth]{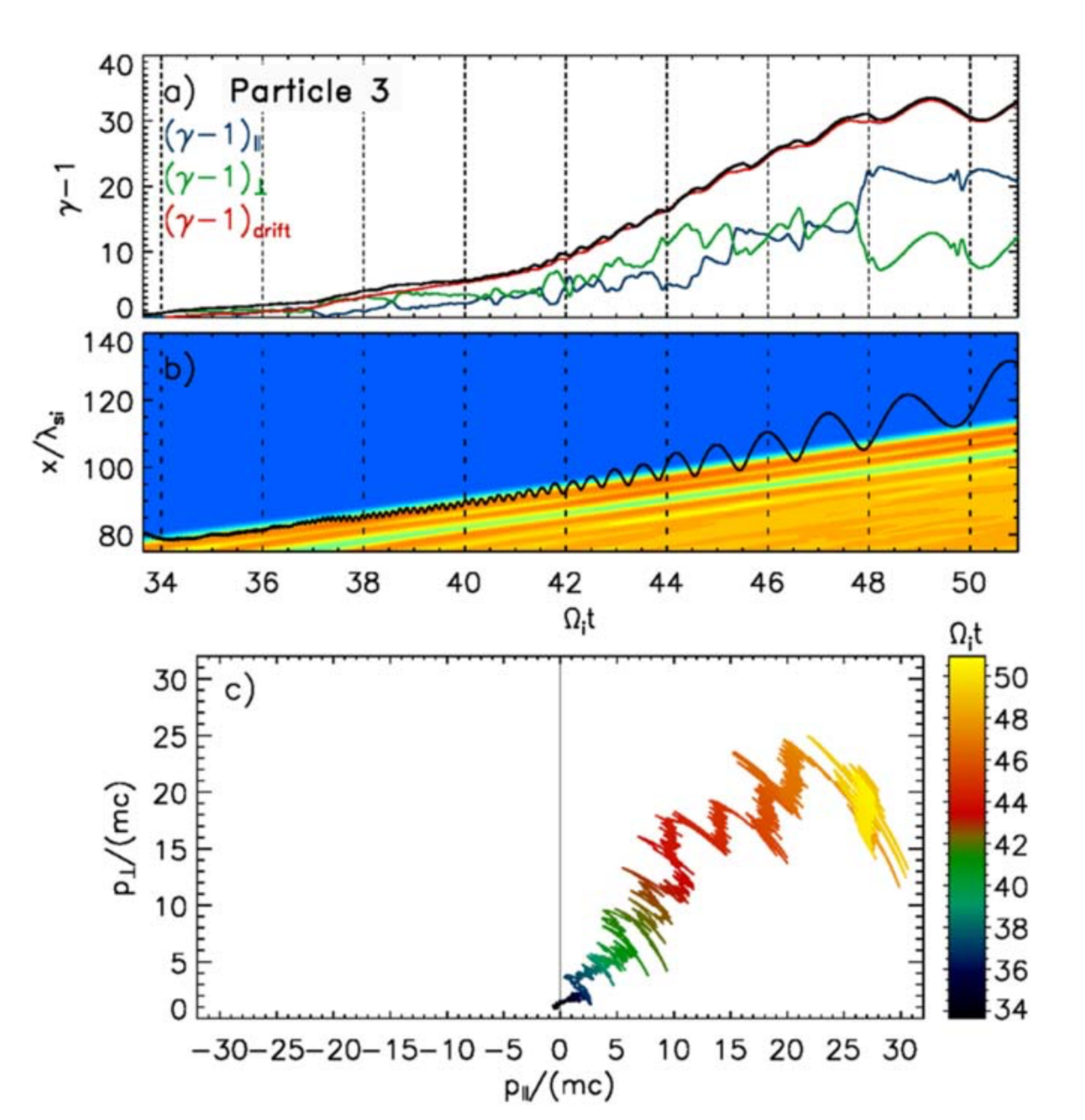}
    \caption{Trajectory of accelerated electron in the 2D simulation by \citet{Kobzar2021}: (a) energy history, (b) particle position $x$ and average density time evolution, (c) particle trajectory in velocity space. This figure is reproduced from \citet{Kobzar2021}.}
    \label{fig:kobzar2021}
\end{figure}

Analysis of energetic electron trajectories was also presented by \citet{Kobzar2021}, in which a low-Mach-number high-$\beta$ quasi-perpendicular shock ($\MA = 6.1, \thetabn = 75\degr, \betai = \betae = 2.5, \wpe/\wce = 4, m_i/m_e = 100$) was considered for the application to merger shocks in galaxy clusters. Although such a shock is substantially different from that of \citet{Matsumoto2017} in character, some energetic electrons exhibit trajectories that are qualitatively consistent with those expected from the SSDA theory. An example is shown in \figref{fig:kobzar2021}. We see that the energization occurs when the particle is interacting with the shock transition layer. It should be noted that the interaction time $\sim 16 \wci^{-1}$ is longer than that expected from the adiabatic theory $\sim \wci^{-1}$ \citep{Krauss-Varban1989b}, indicating a better confinement efficiency within the shock. Indeed, the shock contains a variety of plasma waves, both electrostatic and electromagnetic in nature, in a broad frequency range. In particular, quasi-parallel propagating whistler-mode waves, ion-scale rippling modes, oblique electromagnetic waves likely be generated by the electron firehose instability may have played the role for scattering. A similar result was recently reported by \citet{Ha2021} as well.

While the discussion above yet remains qualitative, SSDA has been found in shocks in substantially different parameter regimes. It is natural to expect that different kinds of plasma waves dominate the shocks with different parameters. Nevertheless, the nature of wave mode is irrelevant in the SSDA theory as long as the phase speed is negligible compared to the particle velocity.

It may be interesting to note also that efficient electron energization at nearly perpendicular shocks has been found by test-particle simulations in the electromagnetic field reproduced by hybrid simulations \citep[e.g.,][]{Lowe2000,Burgess2006,Trotta2019}. In particular, the surface rippling leads to more efficient electron acceleration as it provides additional electron scatterings during the interaction with the shock. While the lack of electron-scale waves in the hybrid simulations does not allow us to make a direct comparison with the theory, there is a certain sense of similarity with SSDA. We expect that high-frequency whistler waves unresolved by the hybrid simulations will enhance the acceleration efficiency at low energies or at less oblique shocks, in which the ion-scale surface fluctuations are unable to strongly scatter electrons.

\subsection{Connection with Diffusive Shock Acceleration}

An injection model must consider how it is connected to the subsequent DSA. As the SSDA theory is formulated under the diffusion approximation, it is possible to unify SSDA and DSA into a single model that consistently describes the electron acceleration below and above the injection threshold energy.

\citet{Amano2022a} showed that the solution of the diffusion-convection equation for an oblique shock of finite thickness can reproduce both SSDA and DSA, depending on the assumption of the scattering rate. If the diffusion length is much longer than the shock thickness $\ldiff/\Ls \gg 1$, the solution reduces to the standard DSA because the shock is essentially seen as a discontinuity by the accelerated particles. The spectral index then becomes $q = - \partial \ln f / \partial \ln p = 3 r / (r - 1)$ where $r$ is the shock compression ratio. On the other hand, it is approximately given by
\begin{align}
    q \approx 3
    \left[
        1 + \left(
            \ldiff \left< \frac{\partial B}{\partial x} \right>
            \right)^{-1}
    \right]
    \approx 3
    \left[
        1 + \left( \frac{\ldiff}{\Ls} \right)^{-1}
    \right]
\end{align}
for SSDA with $\ldiff/\Ls \lesssim 1$ (see, \citet{Amano2022a}). Therefore, by definition, there is a lower limit for the spectral index for SSDA $q \gtrsim 6$, indicating SSDA produces a much steeper spectrum than the canonical DSA in the strong shock limit ($q = 4$). We note that, in general, the spectrum will not be a pure power-law because $\ldiff$ may be an energy-dependent quantity. It may be worth mentioning that particle acceleration at a finite-thickness shock has been known to produce a steeper spectrum. Particle acceleration at cosmic-ray modified shock is a physically motivated example \citep[e.g.,][]{Drury1982}. It is also discussed in the context of numerical modeling of DSA using stochastic differential equations (SDEs) because a finite shock thickness is inevitable in this approach \citep{Strauss2017}. However, that it is relevant to the issue of electron injection has never been appreciated.

\begin{figure}[tbp]
    \centering
    \includegraphics[width=1.00\textwidth]{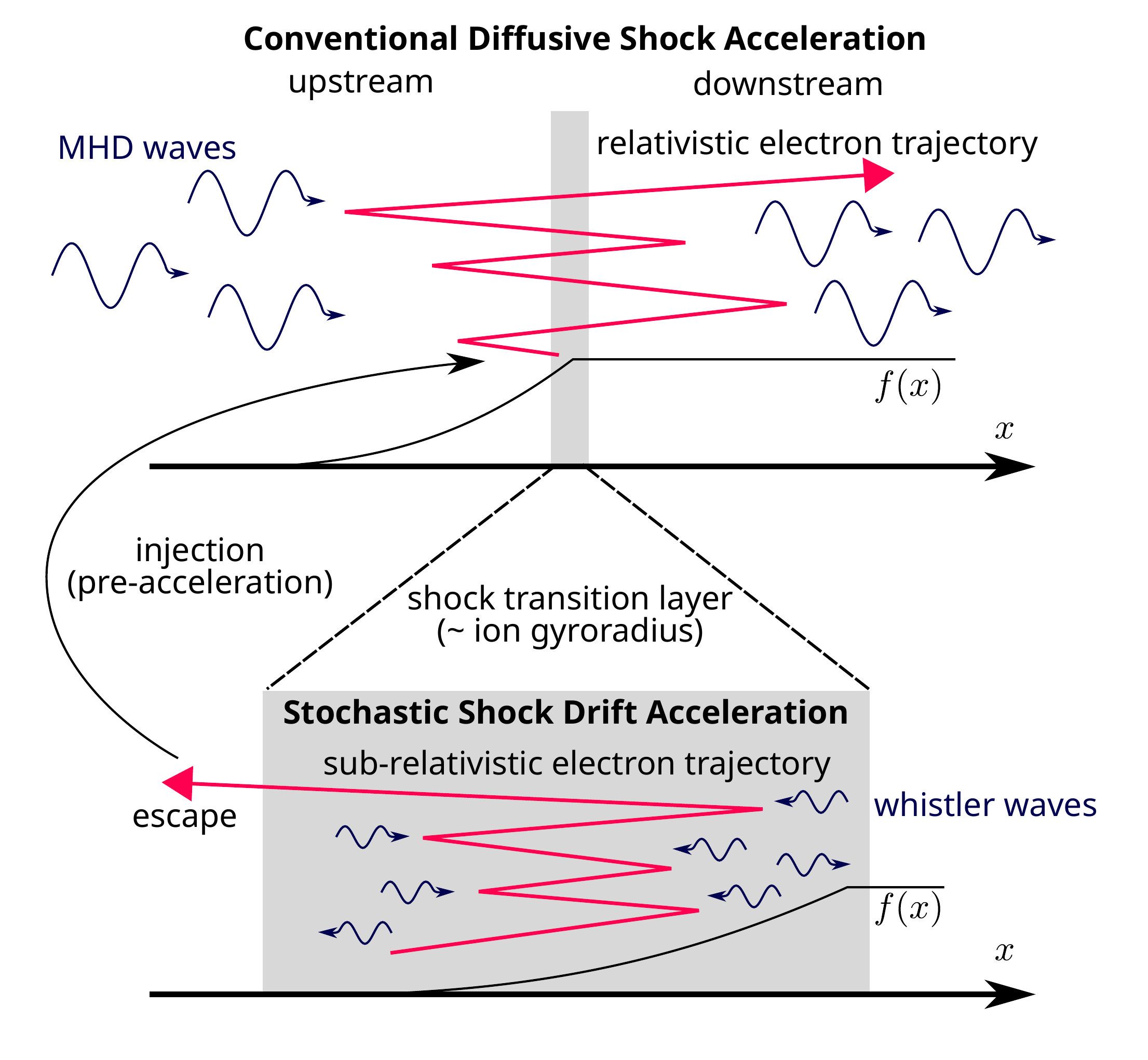}
    \caption{Schematic illustration showing the relation between DSA and SSDA. In contrast to DSA, whereby particles diffuse long distances both in the upstream and downstream, particles accelerated by SSDA are confined within the shock transition layer. Particles accelerated by SSDA will eventually escape away from the shock and may be accelerated further by DSA if the energy is higher than the injection threshold energy. This figure is reproduced from \citet{Amano2020}.}
    \label{fig:amano2020b}
\end{figure}

The injection scheme considered here is schematically illustrated in \figref{fig:amano2020b}. We expect that SSDA will be a dominant process at low energy for which $\ldiff/\Ls \lesssim 1$ is satisfied. As the particle energy increases, the diffusion length tends to increase and at some point becomes larger than the shock thickness, indicating that particles beyond this energy will escape out from the shock transition layer. This thus determines the maximum achievable energy through SSDA. If the escaping particle energy is larger than the minimum energy required for injection, it will be accelerated further by DSA. The resulting energy spectrum will be steep at low energy, where SSDA is dominant. As increasing the energy, it is connected smoothly to the hard standard DSA spectrum beyond the injection threshold. From this discussion, it is clear that the scattering rates both inside and outside the shock transition layer are crucial for injection. Namely, the scattering rate inside the shock determines the maximum momentum achievable through SSDA $p_\textrm{max, SSDA}$ according to \eqref{eq:threshold}, whereas that outside the shock controls the minimum momentum required for DSA $p_\textrm{inj}$ (see, Section \ref{sec:2-electron-injection}). The injection into DSA with the present scheme requires $p_\textrm{max, SSDA} \gtrsim p_\textrm{inj}$.


\begin{figure}[tbp]
    \centering
    \includegraphics[width=0.80\textwidth]{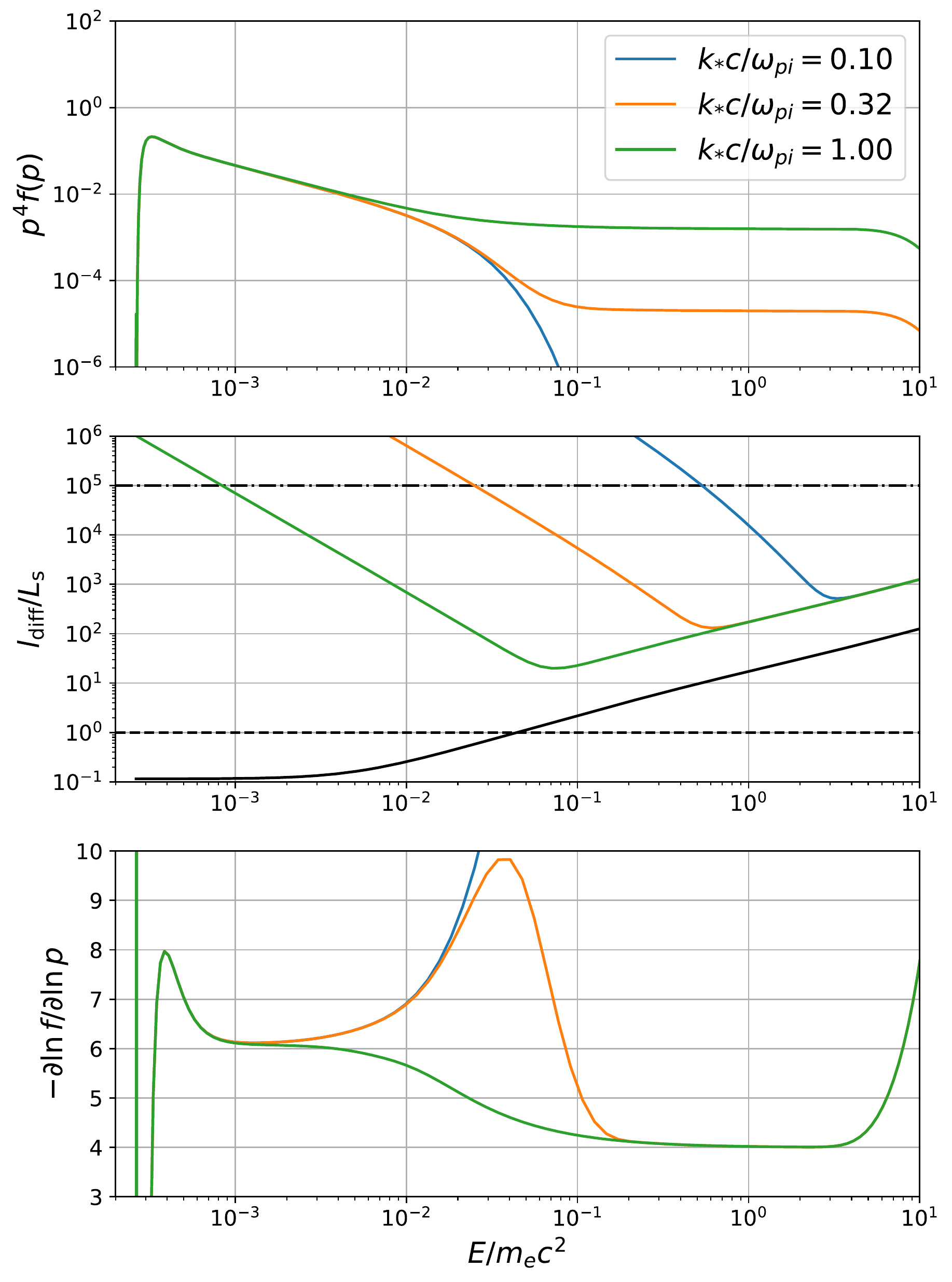}
    \caption{Numerical solutions to the diffusion-convection equation (\ref{eq:diffusion-transport}) for a quasi-perpendicular shock with a finite thickness. Three solutions with the dissipation scale length $k_{*}c/\wpi = 0.10, 0.32, 1.00$ and a fixed $\thetabn = 85\degr$ are shown. The different color indicates different values of $k_{*} c / \wpi$. The top panel shows the energy spectrum. The diffusion length normalized to the shock thickness is shown in the middle panel, where the black and colored lines indicate the diffusion length inside and outside the shock transition layer, respectively. The horizontal dashed and dash-dotted lines indicate the shock thickness and the system size $\Lfeb$ (free escaping boundary), respectively. The bottom panel shows the local spectral slope $-\partial \ln f/\partial \ln p$. This figure is reproduced from \citet{Amano2022a}.}
    \label{fig:amano2022}
\end{figure}

\figref{fig:amano2022} shows numerical solutions obtained by solving the diffusion-convection equation for a smooth shock transition layer. We use the magnetic field profile given by
\begin{align}
	B_z (x) =
	\frac{B_{z,2} - B_{z,1}}{2} \tanh \left( \frac{x}{L_s} \right) +
	\frac{B_{z,2} + B_{z,1}}{2}.
\end{align}
Note that we assume that $B_z$ is the compressional component of the magnetic field. A stationary MHD shock indicates $V_x B_z = \textrm{const}$, which gives the flow profile
\begin{align}
	V_x (x) = V_{x,1} \frac{B_{z,1}}{B_z(x)}.
\end{align}
The subscript 1 and 2 indicate the upstream and downstream quantities, respectively. To represent the finite system size that is important to determine the injection threshold energy, we introduced the free-escaping boundary in the upstream at $\Lfeb$ ahead of the shock. The shock parameters $\MA = 10$, $\thetabn = 85\degr$ are chosen for comparison with typical high-Mach number quasi-perpendicular Earth's bow shocks. For the three solutions shown in \figref{fig:amano2022}, we use the scattering rate within the shock transition layer fixed to a model that mimics the measurement of Earth's bow shock \citep{Amano2020}. The scattering rate outside the shock is not known in general. \citet{Amano2022a} nevertheless adopted a model that is parameterized by the dissipation length scale of turbulence $k_{*} c/\wpi$. It is finite only at sufficiently high energy particles that can interact with long-wavelength fluctuations $k < k_{*}$, but otherwise, it rapidly decreases with decreasing the particle energy. In other words, $k_{*}c/\wpi$ controls the diffusion length in the upstream region, which must be smaller than the system size $\Lfeb$ for DSA. See, \citet{Amano2022a} for more details.

In the top panel, we see that the low-energy spectra are nearly identical in all three cases as the particles in this energy range are confined within the shock $\ldiff/\Ls \lesssim 1$. As we expect, the high-energy part of the spectrum depends crucially on $k_{*} c/\wpi$. For $k_{*} c/\wpi = 0.1$, there appears a clear cutoff at high energy. Looking at the middle panel, we understand that the diffusion length $\ldiff$ (the blue line for $k_{*}c/\wpi = 0.1$) at the cutoff energy ($E/m_e c^2 \sim 3 \times 10^{-3}$) is much longer than the system size $\Lfeb$ (the black dash-dotted line). It is then reasonable that particle acceleration by DSA does not operate. For the case with $k_{*} c/\wpi = 1.0$ shown in green, the spectrum becomes gradually harder as increasing the energy because the diffusion length for this case is much smaller than the system size $\ldiff/\Lfeb \ll 1$ at the transition energy between SSDA and DSA. In the bottom panel, the local spectral slope is shown. The high energy spectral index $q \approx 4$ for $k_{*}c/\wpi = 1.0$ indicates that the injection into DSA is achieved in this case. Note that the steepening at the highest energy seen for $k_{*}c/\wpi = 1.0$ is due to the finite system size and is not due to an intrinsic limit of the model.

We should emphasize that the validation of the diffusion approximation by in-situ observations at Earth's bow shock provided a significant step forward in the theoretical development. This allowed us to construct a model that describes both the injection process by SSDA at low energy and the further energization by DSA at high energy on equal footing. It is important to remind the reader that the electron injection has been a challenging issue in DSA because of the lack of scattering agents. Indeed, the diffusion approximation for low-energy electrons should only be appropriate within the thin layer, which is assumed as a discontinuity in the standard DSA theory. While the overall acceleration scheme is described by the diffusion approximation, there is no need to provide scattering agents except for the close vicinity of the shock, as long as the low-energy electron acceleration completes in the local region.

\subsection{Dependence on Shock Parameters}

The injection scheme under the diffusion approximation discussed so far provides the framework to analyze the issue of electron injection in a quantitative manner. Unfortunately, however, the capability of particle acceleration with both SSDA and DSA depends crucially on the scattering rate. This obviously indicates a need for understanding the wave generation processes and the resulting pitch-angle scattering rate.

Nevertheless, it is instructive to summarize the expected particle acceleration mechanisms for a given efficiency of pitch-angle scattering. To simplify the discussion, the energy range of interest will be limited up to mildly relativistic energy, and corrections due to the Lorentz factor will be ignored in the following. We assume that the scattering rate within the shock transition layer is given by
\begin{align}
    D_{\mu\mu}(p) =
    \begin{cases}
        \Dmax \left( \dfrac{p}{p_{*}} \right)^2 &
        p < p_{*}, \\
        \Dmax &
        p \geq p_{*}.
    \end{cases}
    \label{eq:dmumu_model}
\end{align}
This particular form is phenomenological and motivated by in-situ observations \citep{Amano2020}. If $\Dmax$ is sufficiently large, this model will reproduce the power-law tail of suprathermal electrons observed in "efficient" bow shock crossing events \citep{Gosling1989,Oka2006}. The saturation beyond momentum $p_{*}$ is introduced to keep the scattering rate below the Bohm limit $\Dmax < \wce$. Note that the following discussion is independent of the choice of $p_{*}$ as long as $\Dmax$ is sufficiently large. In-situ observation by \citet{Amano2020} suggests that the corresponding energy is around a few keV (see, the panel (B) in \figref{fig:amano2020a}).

\begin{figure}[tbp]
    \centering
    \includegraphics[width=1.0\textwidth]{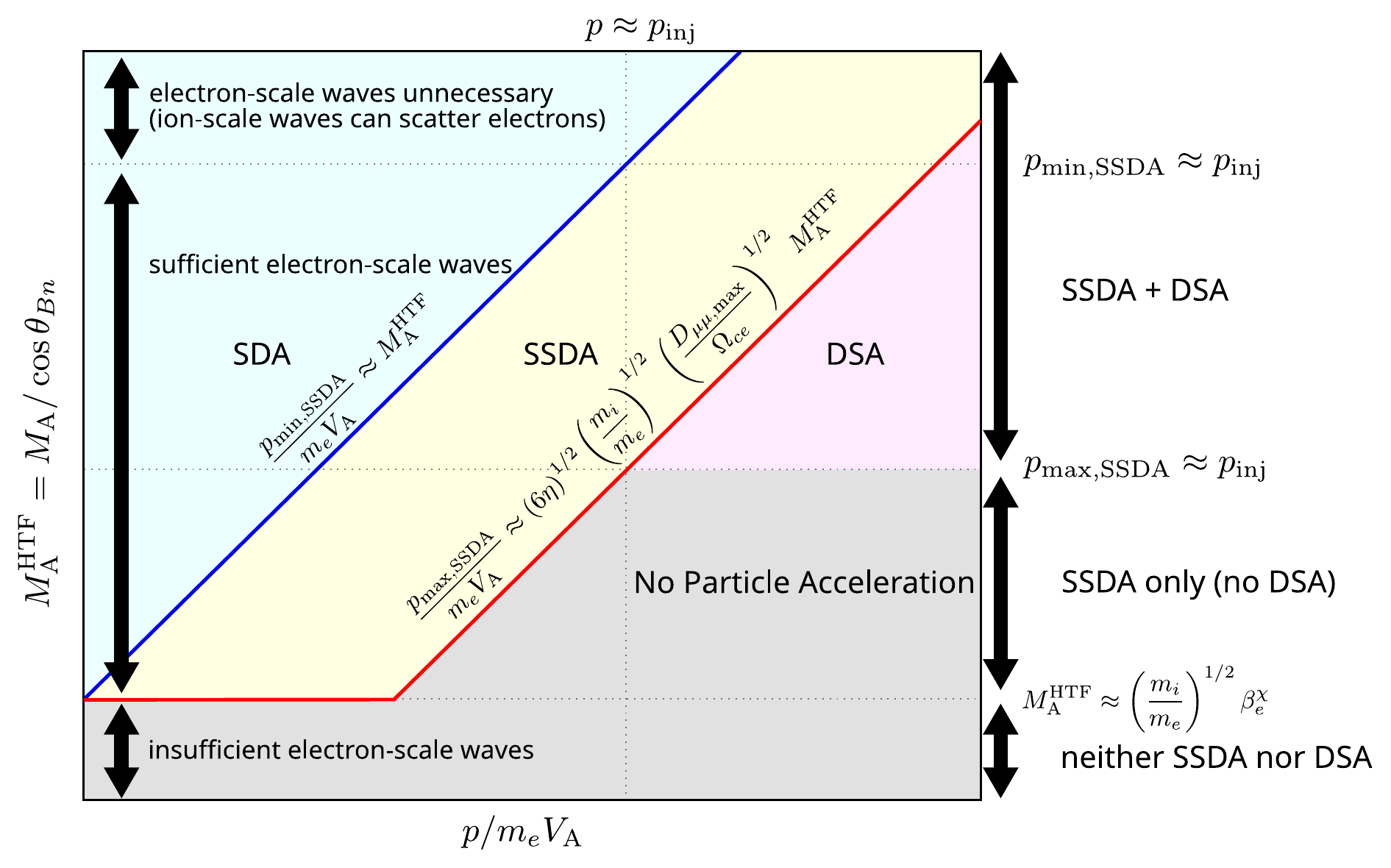}
    \caption{Summary of expected electron acceleration mechanisms. The blue, yellow, and pink areas indicate the parameter ranges where SDA, SSDA, and DSA will take place. No particle acceleration is expected in the gray area. The thick blue and red lines indicate respectively the minimum and maximum energies for SSDA. The diagram assumes the scattering rate given by \eqref{eq:dmumu_model}.}
    \label{fig:injection-diagram}
\end{figure}

Under the assumption, SSDA will apply to the momentum range: $p_\textrm{min, SSDA} \lesssim p \lesssim p_\textrm{max, SSDA}$.
The lower bound in momentum
\begin{align}
    \frac{p_\textrm{min, SSDA}}{m_e \VA} \approx \MAHTF
\end{align}
corresponds to the minimum momentum below which the diffusion approximation becomes invalid \citep{Amano2022a}. It is the same order of magnitude as the momentum gain expected from the adiabatic SDA. Therefore, we may consider that the adiabatic SDA provides the seed population for SSDA. On the other hand, the upper bound
\begin{align}
    \frac{p_\textrm{max, SSDA}}{m_e \VA} \approx
    \left( 6 \eta \right)^{1/2}
    \left( \frac{m_i}{m_e} \right)^{1/2}
    \left( \frac{\Dmax}{\wce} \right)^{1/2}
    \MAHTF.
\end{align}
is determined by \eqref{eq:threshold} and is dependent on $\Dmax$. The particle acceleration to higher energies by DSA may be possible if $p_\textrm{max, SSDA} \gtrsim p_\textrm{inj}$ where $p_\textrm{inj}$ is the injection threshold momentum for DSA. (Otherwise, the spectrum will be cutoff at $p_\textrm{max, SSDA}$.) The injection threshold momentum can be estimated by \eqref{eq:pinj_resonance} for a given dissipation scale length $k_{*}$ \citep{Amano2022a}. It is important to note that the absolute momentum range for SSDA (i.e., both $p_\textrm{min, SSDA}$ and $p_\textrm{max, SSDA}$) linearly scales with the shock speed and is independent of the \Alfven speed. On the other hand, the injection threshold $p_\textrm{inj}$ is proportional to the \Alfven speed (see, \eqref{eq:pinj_resonance}). In other words, the difficulty of injection is dependent on the magnetization parameter $\sigma_s = (\wcs/\wps)^2 \propto (\VA/c)^2$, which often characterizes the collisionless shock dynamics.

The expected particle acceleration mechanisms based on these estimates are summarized in \figref{fig:injection-diagram}. The blue, yellow, and pink regions respectively indicate parameter regimes where the major electron energization mechanisms are due to SDA, SSDA, and DSA. No efficient electron acceleration is expected to occur in the region shown in gray. The SSDA region (yellow) is separated from the others by $p_\textrm{min, SSDA}$ and $p_\textrm{max, SSDA}$ represented by the thick blue and red lines, respectively. The minimum Mach number for the DSA region (pink) is determined by $p_\textrm{max, SSDA} \approx p_\textrm{inj}$, which may be rewritten as
\begin{align}
    \MAHTF \gtrsim
    \left( 6 \eta \right)^{-1/2}
    \left( \frac{m_i}{m_e} \right)^{1/2}
    \left( \frac{\Dmax}{\wce} \right)^{-1/2}
    \left( \frac{k_{*}c}{\wpi} \right)^{-1}
\end{align}
Lower $\MAHTF$ shocks may be able to produce suprathermal electrons within the shock transition layer via SSDA but fail to inject them into DSA because the maximum energy is smaller than the injection threshold. The explicit dependence on $\Dmax$ makes it difficult to predict the exact condition for the shock parameters. However, the theory indicates that the injection will occur preferentially at higher $\MAHTF$ shocks.

Note that \figref{fig:injection-diagram} assumes that the electron acceleration takes place only at shocks satisfying the condition:
\begin{align}
    \MAHTF \gtrsim (m_i/m_e)^{1/2} \betae^{\chi}.
    \label{eq:wave-critical-mach}
\end{align}
Here we have introduced an unknown exponent $\chi$, which may be in the range $-1/2 \leq \chi \leq +1/2$. There has been empirical evidence suggesting that the electron acceleration becomes efficient at whistler-super-critical shocks, which, if we ignore a factor of order unity, may be written as $\MAHTF \gtrsim (m_i/m_e)^{1/2}$ corresponding to $\chi = 0$ \citep{Oka2006}. We should mention that there is a possible dependence on $\betae$ that has not yet been explored with spacecraft observations.

\begin{figure}[tbp]
    \centering
    \includegraphics[width=1.0\textwidth]{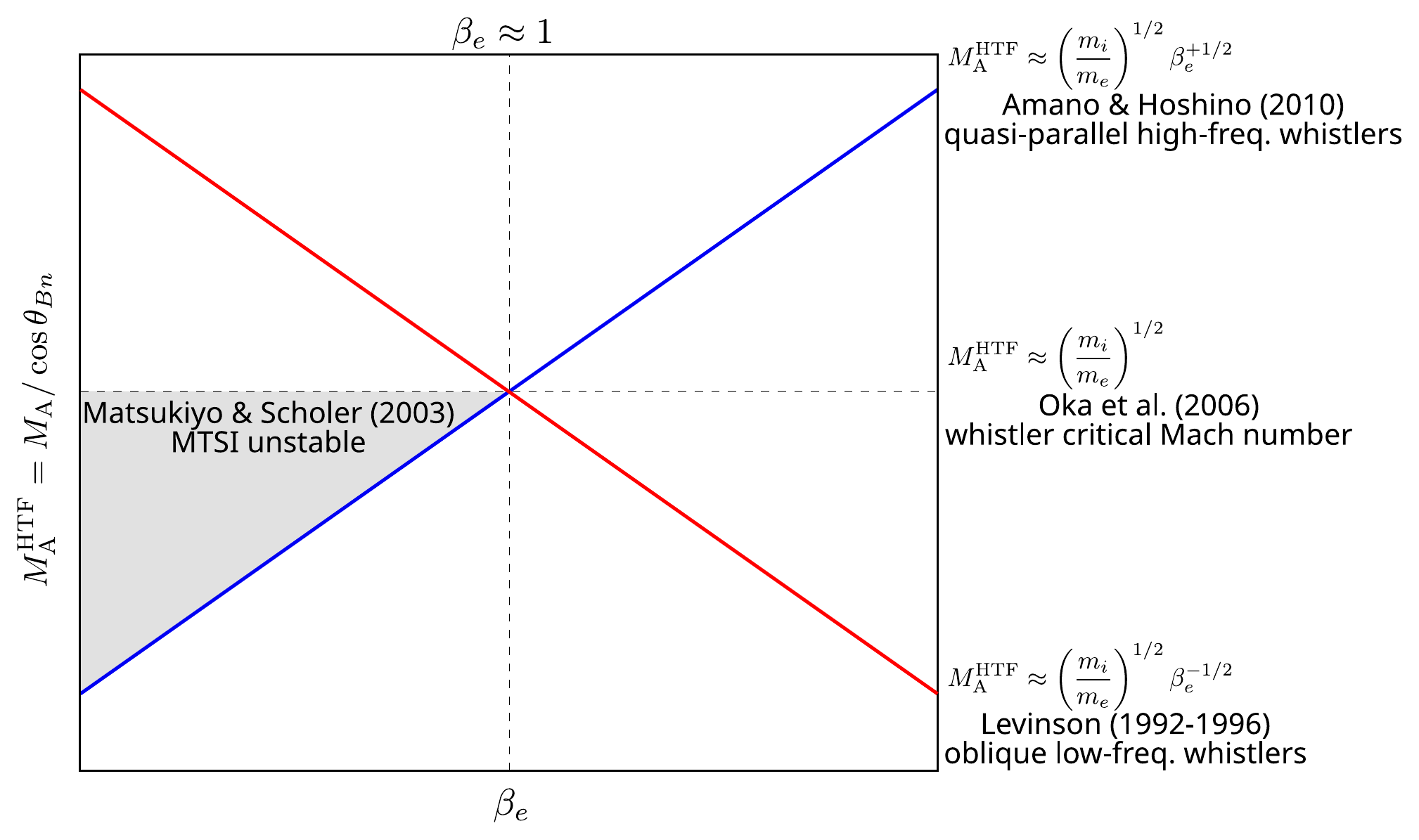}
    \caption{Critical Mach numbers as functions of $\betae$. The blue line represents the critical Mach number for quasi-parallel high-frequency whistler wave excitation given by \citet{Amano2010}. The red line indicates the model proposed by \citet{Levinson1992a,Levinson1994,Levinson1996} that considers the condition for oblique low-frequency whistler wave excitation. The horizontal dashed line indicates the whistler critical Mach number, which was shown to regulate the energetic electron spectrum at Earth's bow shock \citep{Oka2006}. The gray shaded area indicates the region where the system is unstable against MTSI \citep{Matsukiyo2003a}. Note that all these conditions ignore numerical factors of order unity and should be understood as rough estimates.}
    \label{fig:critical-mach-mahtf}
\end{figure}

Interestingly enough, aside from a numerical factor of order unity, the dependence represented by \eqref{eq:wave-critical-mach} arises in different wave-generation mechanisms, which are summarized in \figref{fig:critical-mach-mahtf}. \citet{Levinson1992a,Levinson1994,Levinson1996} first pointed out that low-frequency whistler waves of oblique propagation may be excited by shock-accelerated electrons if $\MAHTF \gtrsim (m_i/m_e) \betae^{-1/2}$, which is shown in red in \figref{fig:critical-mach-mahtf}. Although the waves generated by this mechanism will propagate toward upstream, the oblique propagation nature enables the scattering of upstream-moving electrons via the anomalous cyclotron resonance (see, e.g., \citet{Karimabadi1992} for the roles of higher-order cyclotron harmonic resonances in efficient pitch-angle scatterings by obliquely propagating waves). This may result in the efficient confinement of electrons required for particle acceleration. It should be noted that the mechanism of the resonant electron firehose-like instability observed in kinetic simulations \citep[e.g.,][]{Matsukiyo2011,Guo2014a,Guo2014b,Xu2020,Kobzar2021,Ha2021} appear to be similar to Levinson's mechanism in that the anomalous cyclotron resonance with obliquely-propagating waves destabilizes the waves.  Note also that a similar oblique whistler excitation mechanism driven by an electron heat flux has been discussed recently in different contexts \citep[e.g.,][]{Roberg-Clark2018,Krafft2010,Verscharen2019}.

On the other hand, \citet{Amano2010} considered an electron beam reflected by SDA interacting with the background plasma. The reflected electrons will have a loss-cone-like structure in velocity space, which can be the source of free energy. The system may become unstable when $\MAHTF \gtrsim (m_i/m_e)^{1/2} \betae^{1/2}$, which is shown in blue in \figref{fig:critical-mach-mahtf}. The resulting wave will be a high-frequency quasi-parallel propagating whistler wave. The loss-cone-like velocity distribution function (VDF) allows the generation of waves propagating toward the downstream that can scatter upstream-moving electrons via the normal cyclotron resonance. Again, the waves may play a role in the confinement of electrons in the close vicinity of the shock. We note that both of the above scenarios are self-generation mechanisms in the sense that the shock-accelerated electrons can generate the required waves for particle acceleration by themselves.

Another potentially important mechanism is the modified-two-stream instability (MTSI), which also has a similar dependence. The heating and acceleration of electrons associated with MTSI have been discussed extensively in the literature \citep{Wu1983,Matsukiyo2003a,Matsukiyo2006a}. In the shock transition layer, the MTSI is driven either by the reflected ions or the upstream core ions interacting with the magnetized background electrons. As a result, obliquely propagating low-frequency whistler-mode waves will be generated. \citet{Matsukiyo2003a} showed that $(m_i/m_e)^{1/2} \betae^{+1/2} \lesssim \MAHTF \lesssim (m_i/m_e)^{1/2}$ provides the condition for instability (the gray shaded area in \figref{fig:critical-mach-mahtf}). Note that the lower bound is given by the electron Landau damping stabilization, while the upper bound is required so that the ion beam mode will intersect with the oblique whistler wave dispersion branch. In contrast to the above two self-generation mechanisms, the ion dynamics play an important role in wave generation via MTSI.

In principle, the two self-generation mechanisms can equally be consistent with the observations by \citet{Oka2006} because $\betae \sim 1$ on average in the solar wind. Recall again that our discussion ignores numerical factors of order unity and is rather qualitative. Careful observational data analysis in the future may be able to distinguish the whistler generation mechanisms. We conjecture that the loss-cone driven mechanism by \citet{Amano2010} (and possibly the MTSI if $\MAHTF \lesssim (m_i/m_e)^{1/2}$) will play a role in the wave generation and resulting electron acceleration at $\betae \ll 1$. On the other hand, Levinson's mechanism may favor $\betae \gg 1$. In other words, different mechanisms may become dominant in different plasma environments. Furthermore, both of these mechanisms may potentially play roles at the same time under the average solar wind with $\betae \sim 1$. In any case, it is clear that the electron-driven wave generation mechanisms need to be investigated in the future for more quantitative electron injection modeling.

Finally, we comment that, in principle, the electron-scale waves may not be necessary at extremely strong shocks. In other words, if a single SDA is sufficient to energize electrons beyond the injection threshold, the reflected electrons can be directly injected into DSA \citep{Amano2007}. The condition $p_\textrm{min, SSDA} \gtrsim p_\textrm{inj}$ may be rewritten as follows:
\begin{align}
    \MAHTF \gtrsim
    \left( \frac{m_i}{m_e} \right)
    \left( \frac{k_{*}c}{\wpi} \right)^{-1}.
\end{align}
As we shall see in Section \ref{sec:5}, we anticipate that such a strong shock will be unstable against the Weibel Instability (WI). There will be magnetic turbulence of substantial amplitudes at the ion inertial scale. Therefore, in this regime, the Weibel-generated turbulence will be able to directly scatter the electrons, which may be accelerated to ultra-relativistic energies within the shock transition layer \citep{Matsumoto2017}. Note that, however, the simplified schematic diagram shown in \figref{fig:injection-diagram} may not necessarily be representative in this regime, where the shock may behave much more violently than the conventional magnetized shock.

\clearpage
\section{Shock Surfing Acceleration} \label{sec:4}

\subsection{Acceleration Mechanism} \label{sec:4-mechanism}

Shock surfing acceleration (SSA) has been considered as an efficient particle acceleration mechanism at a quasi-perpendicular shock that may operate both for ions and electrons but in slightly different contexts. In both cases, an electrostatic force perpendicular to the magnetic field that can reflect off the particles coming into the shock is required. For positively charged ions, the cross-shock potential can play the role, and the reflected ions are energized by traveling in the direction of the convection electric field during a reflection \citep{Sagdeev1996}. If the spatial scale of the cross-shock potential is small, on the order of the electron inertial length, the electric field will be very strong such that the reflection occurs multiple times. The multiple reflections lead to efficient particle acceleration, which was considered as a model for pickup ion acceleration at the solar wind termination shock \citep{Lee1996a,Zank1996a}. While the potential scale is currently believed to be much broader, and it may not be such efficient as originally thought, it is worth understanding as essentially the same process may appear in different contexts \citep{Katsouleas1983}.

\begin{figure}[tbp]
    \centering
    \includegraphics[width=1.00\textwidth]{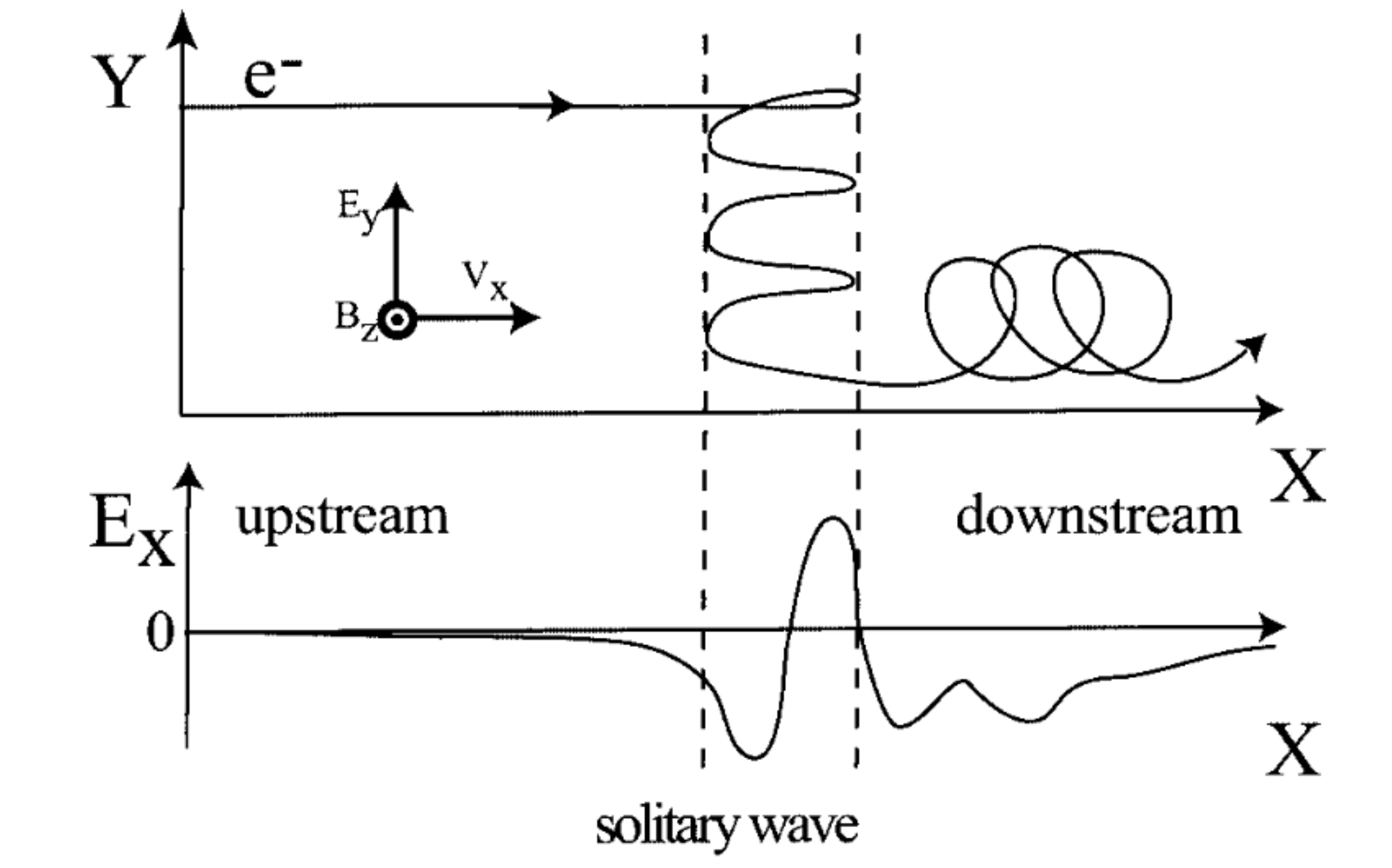}
    \caption{Schematic illustration of SSA for electrons in 1D. The electron moving in the positive $x$ direction is trapped by a solitary electrostatic wave nearly stationary in the shock rest frame. During the trapping, the electron will be accelerated in the negative $y$ direction by the motional electric field arising from the relative velocity difference between the plasma rest frame and the wave propagation. This figure is reproduced from \citet{Hoshino2002}.}
    \label{fig:hoshino2002a}
\end{figure}

Since a negative charge cannot be reflected by the macroscopic cross-shock potential, an electric field of different origins is needed to consider SSA for electrons \citep{McClements2001,Hoshino2002,Schmitz2002}. To see the particle acceleration mechanism quantitatively, let us consider the simplest possible case of a purely perpendicular shock in NIF. We assume that the electromagnetic field in the upstream region is constant and given by $\bm{E} = (E_x, E_y, 0)$ and $\bm{B} = (0, 0, B_z)$. The equation of motion for a non-relativistic electron is given by
\begin{align}
    & \frac{d v_x}{dt} =
    -\frac{e}{m_e} \left( E_x + \frac{v_y}{c} B_z \right),
    \\
    & \frac{d v_y}{dt} =
    -\frac{e}{m_e} \left( E_y - \frac{v_x}{c} B_z \right).
\end{align}
There is a finite motional electric field $E_y = \Us B_z/c$ in this frame due to the background plasma flow $\Us$ in the $x$ direction. In the absence of $E_x$ (i.e., in the far upstream), the particle simply performs the $\bm{E} \times \bm{B}$ drift in the $x$ direction. Here we assume the presence of an electrostatic wave with a finite $E_x \propto \cos(k x)$ that is stationary in NIF (i.e., propagating with respect to the background plasma). In this case, the $\bm{E} \times \bm{B}$ drift motion in the $x$ direction is prohibited if $v_y/c \leq \lvert E_x/B_z \rvert$, which will make $v_x$ oscillatory around $v_x \sim 0$. Therefore, the average velocity in the $x$ direction becomes negligble $\left< v_x \right> \sim 0$ and the particle is accelerated in the $y$ direction by the motional electric field $E_y$. The particle acceleration mechanism by SSA is schematically illustrated in \figref{fig:hoshino2002a}, in which a solitary (rather than sinusoidal) wave is considered as suggested by kinetic simulations. The energy gain timescale may be roughly estimated as follows
\begin{align}
    \frac{1}{\tau_{\rm acc, SSA}} =
    \frac{1}{\epsilon} \frac{d \epsilon}{d t} \sim
    \frac{-e v E}{m_e v^2 / 2} \sim
    2 \wce \left( \frac{\Us}{v} \right).
    \label{eq:tacc_ssa}
\end{align}
For a particle with the initial velocity comparable to the upstream flow velocity (i.e., a cold upstream electron), the acceleration timescale for SSA is roughly $\tau_{\rm acc, SSA} \sim O(\wce^{-1})$. This particle acceleration mechanism is qualitatively similar to SDA in that the energy gain comes from the motional electric field. In SDA, it is the gradient-B drift speed with which the particle moves along the electric field. On the other hand, the particle accelerated by SSA moves with its individual velocity along the electric field. This makes the acceleration time by SSA much faster than SDA: $\tau_{\rm acc, SDA}/\tau_{\rm acc, SSA} \sim O(m_i/m_e) \gg 1$.

The maximum achievable energy is estimated by the condition $v_y/c \sim \lvert E_x/B_z \rvert$, which may be rewritten as
\begin{align}
    \frac{E_{\rm max, SSA}}{m_e c^2} \sim
    \frac{1}{2} \left( \frac{E_x}{B_z} \right)^2
\end{align}
for $\lvert E_x/B_z \rvert < 1$. On the other hand, if $\lvert E_x/B_z \rvert > 1$, the condition $v_y/c \leq \lvert E_x/B_z \rvert$ will always be satisfied even for a particle with $v_y \sim c$. In other words, the reflection by the electric force always occurs no matter how large the particle energy is. In this case, there is no upper limit of the particle energy and is referred to as the unlimited acceleration regime \citep{Katsouleas1983,Hoshino2002}.

It is instructive to look at the same process also in the plasma rest frame where the motional electric field vanishes. In this frame, it is obviously the wave electric field $E_x$ that accelerates the electrons pushed by the propagating wave. In the absence of the background magnetic field, the electrons gain energy by a specular reflection by the wave. A finite magnetic field transfers the particle momentum gained in the $x$ direction to the $y$ direction via the Lorentz force, which makes the resulting energization more efficient. If the wavelength is much longer than the accelerated electron gyroradius (which is, however, not the case for the application to collisionless shocks), it will be accelerated up to the $\bm{E} \times \bm{B}$ drift speed in the $y$ direction. It is known that the electric field vanishes in the frame moving with the $\bm{E} \times \bm{B}$ drift, in which the particle performs a simple gyromotion. The unlimited acceleration condition $\lvert E_x/B_z \lvert > 1$ indicates that there is no such frame where the electric field vanishes. The Lorentz invariance of $\bm{E}^2 - \bm{B}^2$ implies that one can instead find a frame where the magnetic field vanishes. Therefore, the particle no longer performs the gyromotion, and its energy may increase without bounds. Although we will consistently state that the motional electric field accelerates particles in SSA, we should always keep in mind the frame dependence of the problem.

We have seen that a strong electrostatic wave propagating perpendicular to the magnetic field may accelerate electrons to high energies in a short period of time through SSA. It is important to remind the reader that not only the amplitude of the wave but also a finite propagation speed with respect to the plasma is crucial for particle acceleration. The motional electric field that accelerates the particle in the transverse direction arises due to the relative velocity between the plasma and the wave. As we see below, waves with the above-mentioned properties may be generated naturally at the leading edge of the transition layer of very high Mach number shocks.

\subsection{Buneman Instability} \label{sec:4-buneman}


One of the most violent instabilities driven by the reflected ions is the so-called Buneman instability (BI), which is an electrostatic instability and occurs due to the resonance between the ion beam mode and the Langmuir wave (or the upper-hybrid wave in a magnetized plasma). The possible importance of BI in the context of electron energization at collisionless shocks has been recognized for a long time \citep{Papadopoulos1988a,Cargill1988a,Shimada2000,Dieckmann2000a}. The instability requires that the relative streaming velocity between the ions and electrons is greater than the electron thermal velocity because otherwise, the wave will suffer strong electron Landau damping \citep{Matsukiyo2003a}. If we consider the application to collisionless shocks, we obtain the following condition for the linear instability:
\begin{align}
    \MA \gtrsim \frac{1 + \alpha_\textrm{r}}{2}
    \left( \frac{m_i}{m_e} \right)^{1/2}  \betae^{1/2},
\end{align}
where $\alpha_\textrm{r}$ is the fraction of the reflected ions, which is typically $\alpha_\textrm{r} \sim 0.2$ \citep{Matsumoto2012}. We see that the required Mach number is much higher than typical values at planetary bow shocks in the heliosphere. Nevertheless, very high Mach number shocks such as those of young SNRs may satisfy the condition.

The linear instability threshold above should be considered as a necessary condition for SSA. Since the critical parameter for SSA is the maximum wave electric field normalized to the background magnetic field (i.e., $E_x/B_z$), we need to understand the saturation level of the instability to estimate the particle acceleration efficiency. In general, it is a difficult task to obtain a quantitative estimate for the nonlinear saturation level of a kinetic instability. If the amplitude becomes sufficiently large, particle trajectories will substantially be modified from their unperturbed ones. Ultimately, they may be trapped by the wave electric field and start to bounce in the wave potential. One of the classical ways to estimate the saturation level is to equate the trapped-particle bounce frequency and the linear growth rate. By also taking into account the nonlinear mode coupling (or wave steepening) effect, \citet{Ishihara1981} obtained an estimate for the saturation level, which, in our notation, may be written as $E_x/B_z \approx 2 (1 + \alpha_\textrm{r})^{-1} \MA (m_i/m_e)^{-2/3}$. While the \Alfven Mach number is certainly an important parameter, we have to take into account the dependence on the ion-to-electron mass ratio as well.

The discussion so far has been based on the 1D assumption. However, the multidimensional effect is known to have a non-negligible impact. It has been known that BI in the cold electron limit produces a lot of oblique modes with respect to the beam direction \citep{Lampe1974,Dieckmann2006d}. The presence of oblique modes increases the region in velocity space where the resonant wave-particle interaction takes place, which allows the resonant trapping by the waves to start to play a role earlier than in 1D. As a result, the saturation level reduces by a factor of $\sim 1/4$ \citep{Amano2009b}. By taking the multidimensional effect into account, we may rewrite the unlimited acceleration condition $E_x/B_z \gtrsim 1$ as follows \citep{Matsumoto2012}:
\begin{align}
    \MA \gtrsim
    \left( 1 + \alpha_\textrm{r} \right)
    \left( \frac{m_i}{m_e} \right)^{2/3}.
    \label{eq:unlimited-acc}
\end{align}
It is important to recognize that a reduced mass ratio often adopted in PIC simulations directly affects the required \Alfven Mach number. The validity of the above condition has been confirmed in 2D PIC simulations studies \citep{Matsumoto2012,Matsumoto2013,Bohdan2017}.

\subsection{Particle Acceleration in 1D} \label{sec:4-1d}

The nonlinear development of BI results in the generation of a large-amplitude electric field that is associated with a phase-space electron hole in velocity space \citep{Davidson1970a}. The electron hole is generated because the majority of electrons are reflected by the wave electrostatic potential. Such large-amplitude waves may trap a small number of electrons inside the potential structure. As the waves are driven by the reflected ion beam, they propagate in the plasma with a speed comparable to the reflected ion beam. In other words, the wave electrostatic potential structure moving with respect to the background plasma with a speed comparable to the shock speed. Therefore, an electron trapped inside the moving potential structure may be accelerated by a finite motional electric field seen in the wave rest frame (see, \figref{fig:hoshino2002a}).

\begin{figure}[tbp]
    \centering
    \includegraphics[width=1.00\textwidth]{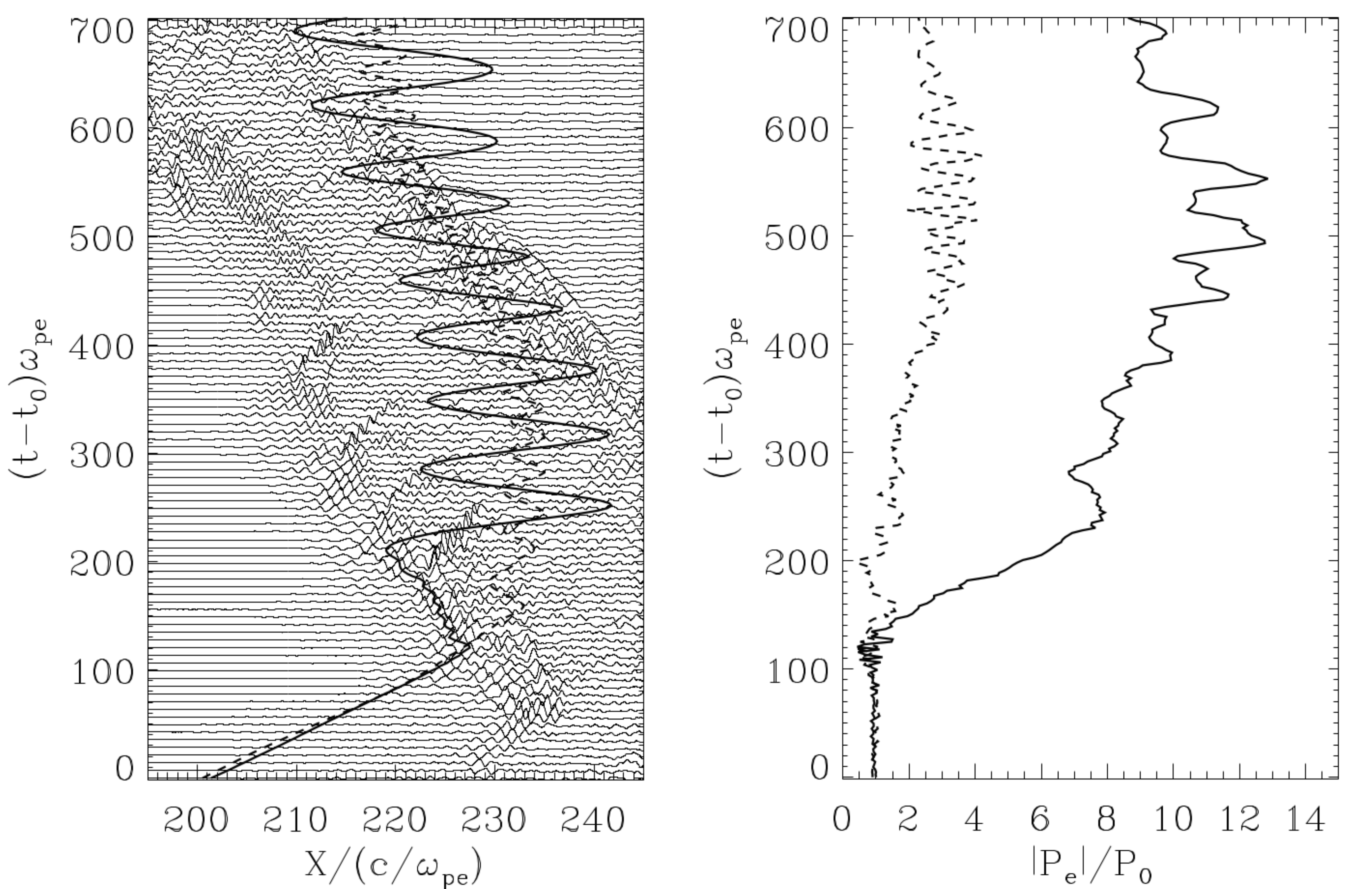}
    \caption{Electron trajectories as seen in the 1D simulation by \citet{Hoshino2002}. The solid and dashed lines represent accelerated and non-accelerated particles, respectively. The enlarged view around the shock transition layer is shown. The left panel indicates the electron positions with the stacked profiles of the electrostatic field $E_x$. The right panel shows the particle momenta. This figure is reproduced from \citet{Hoshino2002}.}
    \label{fig:hoshino2002b}
\end{figure}

Early 1D PIC simulations \citep{McClements2001,Hoshino2002,Schmitz2002} confirmed that electrons with relativistic energies are produced indeed with this mechanism. \figref{fig:hoshino2002b} shows trajectories of characteristic accelerated and non-accelerated electrons obtained in the 1D PIC simulation by \citet{Hoshino2002}. The parameters for simulation are: $\MA = 32$, $\betai = \betae = 0.01$, $\wpe/\wce = 19$, $m_i/m_e = 20$. In the left panel, stacked profiles of the electric field $E_x$ are shown to represent the spatiotemporal evolution. Note that the spatial scale shown is roughly comparable to the shock transition layer, and we see that the shock is propagating in the negative $x$ direction. The solid and dashed lines represent the trajectories of accelerated and non-accelerated particles. The right panel shows the time history of momentum. It is clear that during the efficient energy gain identified in the right panel, the accelerated particle moves to the left approximately with the shock propagation speed, indicating the trapping by a strong electrostatic potential structure also moving with the shock itself. In clear contrast, the non-accelerated particle simply goes through the potential structure and suffers only a limited energy gain through the interaction with the shock. From the simulation result, we find the timescale of energy gain is $\wpe t \sim 100$ or $\wce t \sim 5$, which is roughly consistent with the estimate of \eqref{eq:tacc_ssa}.


\subsection{Particle Acceleration in 2D and 3D} \label{sec:4-multid}

While SSA was shown to be a rapid and very efficient mechanism in producing relativistic electrons in 1D simulations, the validity of its ideal assumption of homogeneity in the transverse direction had been questioned. If there is finite inhomogeneity, the accelerated particles will quickly escape out from the electrostatic potential structure, which will result in the reduction of particle acceleration efficiency. Indeed, the presence of unstable oblique modes expected from the linear theory of BI \citep{Lampe1974} suggests that the electron trapping will never be perfect. By performing 2D PIC simulations in a periodic box, \citet{Ohira2007} concluded that SSA is absent in multidimensions. Similar results have also been reported by \citet{Dieckmann2006d}. However, it was later found that the electron acceleration in a simulation model containing the entire shock structure is surprisingly more efficient than in a periodic model, which deals with only a local portion of the shock.

\begin{figure}[tbp]
    \centering
    \includegraphics[width=1.00\textwidth]{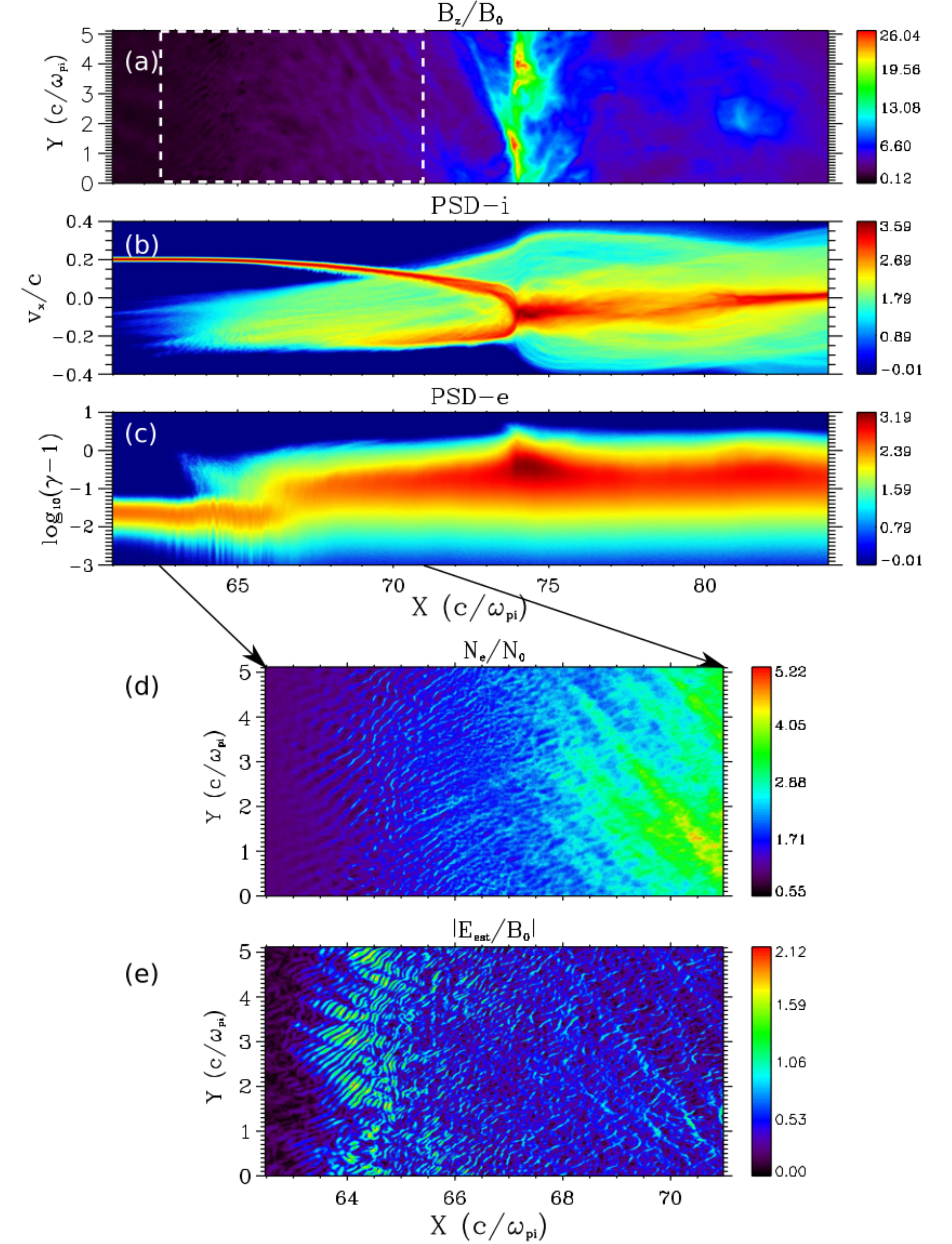}
    \caption{Overview of the 2D PIC simulation by \citet{Matsumoto2012}: (a) Magnetic field profile, (b) phase space density for ions in $x-v_x$, (c) phase space density for electrons in $x-(\gamma-1)$, (d) density profile, (e) electrostatic field amplitudes. The panels (d) and (e) show the enlarged views around the region where the electrostatic waves generated by BI are most intense. This figure is reproduced from \citet{Matsumoto2012}.}
    \label{fig:matsumoto2012}
\end{figure}

\citet{Amano2009a} showed that the nonthermal electron acceleration is quite efficient in a 2D PIC simulation of a Buneman-unstable high Mach number perpendicular shock with the ambient magnetic field purely perpendicular to the simulation plane (i.e., out-of-plane configuration). Although the simulation used an artificial mass ratio of $m_i/m_e = 25$, the same conclusion was later obtained with larger mass ratios as long as the unlimited acceleration condition given by \eqref{eq:unlimited-acc} is satisfied \citep{Matsumoto2012,Matsumoto2013}. The reason for the high electron acceleration efficiency is closely related to the electrostatic wave property generated by BI. \figref{fig:matsumoto2012} shows the 2D simulation results discussed in \citet{Matsumoto2012}: (a) magnetic field profile, (b) ion phase-space diagram, (c) electron phase-space diagram, (d) electron density, (e) electrostatic wave amplitude. The simulation parameters are as follows: $\MA = 30$, $\betai = \betae = 0.5$, $\wpe/\wce = 10$, $m_i/m_e = 100$.

It is clear that the electrostatic waves seen in panel (e) have wavefront oblique with respect to the shock normal (i.e., $x$ direction). The oblique wavefront is due to the gyromotion of reflected ions in the simulation plane, which is perpendicular to the background magnetic field. In the leading edge of the shock, the reflected ions have finite velocities transverse to the shock (the $y$ direction in this case). The simulation results show that the average reflected ion beam velocity is inclined by roughly $45\degr$ from the shock normal direction \citep[not shown, see, e.g.,][]{Matsumoto2012}. The linear theory predicts that the maximum wave growth of BI occurs at the wavenumber parallel to the relative streaming direction. Therefore, the oblique wavefront, which is perpendicular to the ion beam, is a natural consequence of the linear instability.

Another point is that the intense waves are seen only at the leading edge of the shock transition layer with a finite spatial extent. The wave amplitude in this region is kept nearly constant in time. (Note that, however, there is a slower periodic intensification of waves that is associated with the self-reformation of the shock front.) The reason for this is that the free energy is continuously fed into the system in the realistic shock. In other words, the reflected ions always interact with the cold ions and electrons freshly provided by the upstream flow. The situation is completely different from the periodic model in which the free energy is provided only as the initial condition. It is thus natural that the wave intensity will decay after the saturation of the initial wave growth.

Analyses of accelerated electron trajectories found that, in contrast to the original assumption of SSA in 1D, they are not trapped in the wave structures. They move rather freely across different waves in a stochastic manner. Therefore, the inhomogeneity in the transverse direction (i.e., parallel to the wavefront) is not necessarily a limiting factor for particle acceleration. Nevertheless, the unlimited electron acceleration condition is yet an important criterion, which should now be written as $\lvert E_{\rm es}/B_z \rvert > 1$ with $E_{\rm es}$ being the wave electric field along the wave propagation direction. An energetic electron interacting with a wave satisfying the condition may be scattered in gyrophase. This allows some of the electrons to travel along the wavefront longer than in the absence of the scattering. Unless this condition is satisfied, stochastic gyrophase scattering of relativistic electrons by the large-amplitude electrostatic waves will not be realized. Note that the stochastic scattering signature was also found in 2D simulations performed with a periodic box, but the resulting energy spectrum is much softer than in the realistic shock simulation model \citep{Amano2009b}. This is understood partly due to the limited time duration in which the unlimited electron acceleration condition is satisfied.

Since the accelerated particles travel preferentially along the oblique wavefront, they may eventually escape away from the intense wave region that has a finite spatial extent. Since BI is most active in the region where the fresh upstream plasma first interacts with the reflected ions, the intense waves are localized at the leading edge of the shock. Furthermore, it is easy to confirm that the escape direction projected onto the shock normal is always toward upstream. As a result, energetic electrons may be ejected out into the nearly unperturbed upstream region, where the motional electric field induced by the pristine upstream plasma flow can provide a further energy gain. It is interesting to note that the process is very similar to SSA for ions in the sense that the shock reflects the incoming particles, which are then accelerated by the upstream motional electric field \citep{Lee1996a,Zank1996a}. In contrast to SSA for ions, however, the reflection is not due to the quasi-static macroscopic electrostatic potential. The large-amplitude electrostatic waves with the oblique wavefront act as a wall to provide the electron reflection. Clearly, the particle acceleration mechanism is different from the original one discussed based on 1D simulation results. However, we call the process as SSA because of the similarity with the mechanism proposed originally for ions.

\begin{figure}[tbp]
    \centering
    \includegraphics[width=1.00\textwidth]{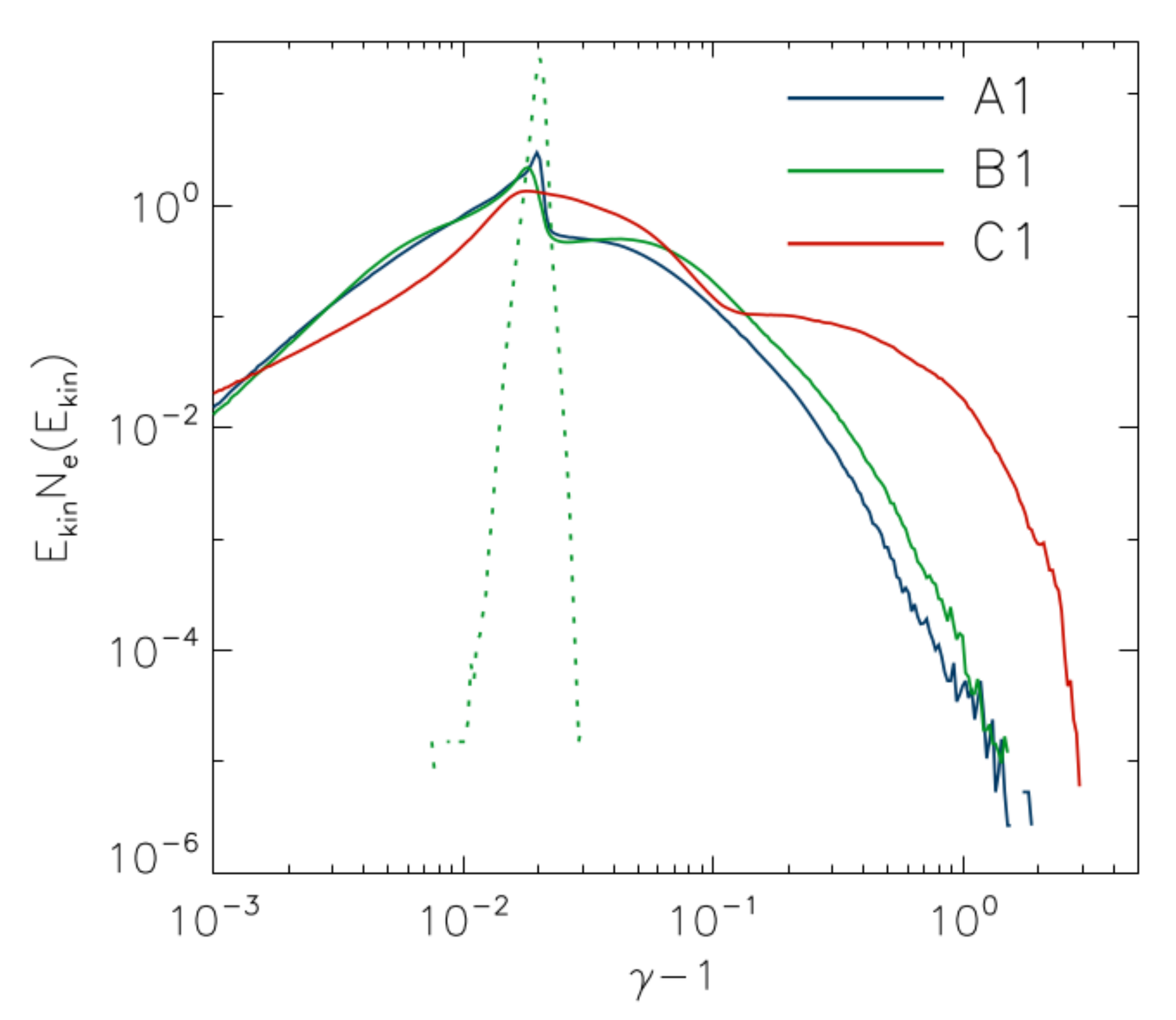}
    \caption{Comparison of energy spectra obtained with different magnetic field orientations with respect to the 2D simulation plane. The spectra are taken in the region where BI is most active. The runs A1, B1, C1 respectively show results for $\psi = 0\degr, 45\degr, 90\degr$, where $\psi$ is the angle between the background magnetic field and the simulation plane. This figure is reproduced from \citet{Bohdan2017}.}
    \label{fig:bohdan2019a}
\end{figure}

The above argument was supported by further numerical experiments performed more recently by \citet{Bohdan2017,Bohdan2019a,Bohdan2019b}. \figref{fig:bohdan2019a} compares the electron energy spectra measured in the Buneman-active region for three different simulation runs performed with the identical setup except for the magnetic field inclination angle $\psi$ with respect to the 2D simulation plane: (A1) $\psi = 0\degr$ (in-plane configuration), (B1) $\psi = 45\degr$, (C1) $\psi = 90\degr$ (out-of-plane configuration). It is clear that the most efficient electron acceleration is seen for the out-of-plane configuration. It is partly due to the largest free energy available for the wave excitation in the out-of-plane, in which the entire relative streaming velocity between the reflected ions and the background plasma is in the simulation plane. In other words, the electrostatic waves will be most intense in the out-of-plane case. Nevertheless, the difference in efficiency in the nonthermal electron production yet remains even if we compare results with similar wave intensities \citep{Bohdan2019a}. These simulation results suggest that the oblique wavefront seen only in the out-of-plane configuration is the most crucial factor. It should be noted that if the finite potential scale in the particle acceleration direction were the limiting factor, the particle acceleration efficiency should have been much better in the in-plane configuration. The simulation result is the complete opposite of this naive expectation.


\begin{figure}[tbp]
    \centering
    \includegraphics[width=1.00\textwidth]{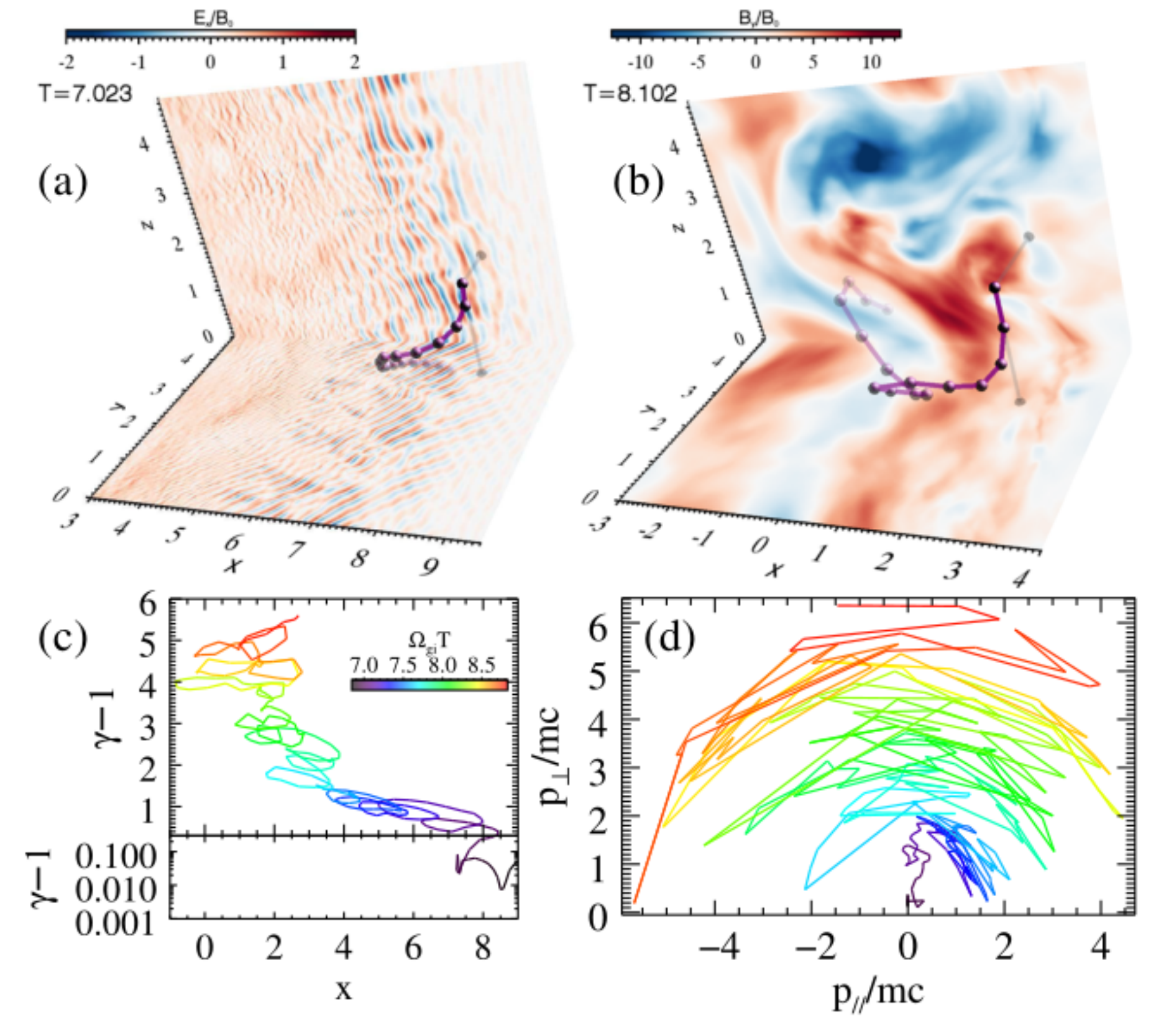}
    \caption{Trajectory of accelerated particle seen in the 3D simulation by \citet{Matsumoto2017}: (a) The time interval for the interaction with Buneman-generated electrostatic waves with the electric field $E_x$ profiles, (b) time interval for the interaction with Weibel-generated magnetic turbulence with the magnetic field $B_y$ profiles, (c) particle energy as a function of position $x$, (d) trajectory in velocity space. The color in the panels (c) and (d) indicates time. Note that $x \sim 8$ indicates the leading edge of the shock transition layer where Buneman-generated waves are most intense.}
    \label{fig:matsumoto2017b}
\end{figure}

The discrepancy in the efficiency of SSA observed in 2D simulations with different magnetic field configurations (but with the same physical setup) clearly indicates a need for fully 3D numerical experiments. While 3D PIC simulations of collisionless shocks have been limited because of the extreme computational resources required, \citet{Matsumoto2017} confirmed that SSA in 3D behaves like in the case seen in the out-of-plane configuration. \figref{fig:matsumoto2017b} shows the 3D PIC simulation results presented in \citet{Matsumoto2017}: (a) electric field, (b) magnetic field, (c) particle energy ($\gamma$ is the Lorentz factor) history as a function of position $x$, (d) trajectory in momentum space $p_{\parallel} - p_{\perp}$. The trajectory of the same particle for specific time intervals are shown in panel (a) and (b) as well. The electrostatic waves have a wavefront inclined roughly $45\degr$ from the $x$ axis in the $x-y$ plane, which is very similar to the 2D simulation result shown in \figref{fig:matsumoto2012} (e). The initially non-relativistic electron enters into the Buneman-active region (i.e., $x \sim 8$) and is energized to mildly relativistic energy in a short period of time. The nearly vertical motion in panel (d) corresponds to the energy gain during this interval. Note that the difference in color represents the time. This efficient particle acceleration is understood as SSA that was found in the 2D out-of-plane configuration. Note also that after the rapid energy gain by SSA, the electron is accelerated even further in the deeper shock transition layer but more slowly than SSA. The slower particle acceleration is understood by SSDA discussed in Section \ref{sec:3}. A more detailed discussion on this will be given in the next subsection.

\subsection{Roles in Electron Injection} \label{sec:4-roles}

Our discussion on the electron acceleration efficiency through SSA has been qualitative so far. Let us now discuss its capability for electron injection. According to the estimate for the injection threshold energy in Section \ref{sec:2-electron-injection}, the maximum achievable energy obtained by SSA should be mildly relativistic or higher. In the original SSA proposed based on the 1D assumption, the unlimited acceleration condition guarantees that electrons trapped within the electrostatic potential can be accelerated to relativistic energies. However, since SSA in 2D and 3D accelerates electrons without trapping, we need a different estimate for the maximum energy.

The energy gain by SSA comes from the motional electric field, as seen by the particles moving approximately with the electrostatic waves. The particles gain energy by traveling in the opposite direction to the motional electric field. The linear thoery suggests that the waves propagate with the reflected ion beam velocity $V_r$ in the upstream rest frame. The travel distance along the electric field is estimated roughly by one gyroradius of particles $r_g \sim v / \wce$ calculated with the velocity $v$ after the energization. Assuming that the particle energy was initially small before the energization, we obtain the condition:
\begin{align}
    \frac{1}{2} m_e v^2 \sim
    e \left( \frac{V_r}{c} B_0 \right) \frac{v}{\wce}.
\end{align}
It is easy to confirm that the particle velocity becomes $v \sim 2 V_r$, which means that the interaction with the turbulent electrostatic waves is simply understood as a specular reflection by a wall moving with the reflected ions. We note that a more detailed estimate with a close comparison with simulation results obtains the same scaling \citep{Amano2009a}. If we use the simulation results $V_r/\Us \sim 3$ \citep{Matsumoto2012,Matsumoto2013,Bohdan2017,Bohdan2019a}, we obtain the energy gain by $E/\Eenif \sim 40$ where $\Eenif = m_e \Us^2/2$ is the upstream electron flow kinetic energy in NIF. Although this estimate and the PIC simulation results indicate that the electron energy becomes relativistic via SSA if the shock speed is high $\Us/c \gtrsim 0.1$, the expected energy in unit of the rest mass energy $m_e c^2$ will be much smaller for typical shock speeds in young SNR shocks with $\Us/c \sim 0.01$. Furthermore, if there is an efficient energy transfer process from ions to electrons, the expected energy gain through SSA may become subdominant in the realistic mass ratio \citep{Bohdan2019b}.


\citet{Amano2009a} suggested that, although the energy gain by a single SSA is limited as estimated above, more efficient particle acceleration might be possible if the same process repeats multiple times. This is exactly the same argument that had been considered for the ion SSA \citep{Zank1996a}. To the best of the authors' knowledge, there has been no evidence for multiple electron SSAs in published PIC simulation results. This might suggest that multiple SSAs are intrinsically prohibited for electrons. Nevertheless, it is also possible that a single SSA is sufficient to energize electrons up to relativistic energies since the shock speed adopted in a typical PIC simulation is unrealistically high. Once the electrons become relativistic $v \sim c$, the particle acceleration will become inefficient as they need to interact with strong electric fields $\lvert E_{\rm es}/B_z \rvert > 1$ for scattering. If this is the case, multiple SSAs may be realized in simulations with more realistic shock speeds $\Us/c \sim 0.01$ with keeping $\MA$ above the unlimited acceleration condition \eqref{eq:unlimited-acc}.

\begin{figure}[tbp]
    \centering
    \includegraphics[width=1.00\textwidth]{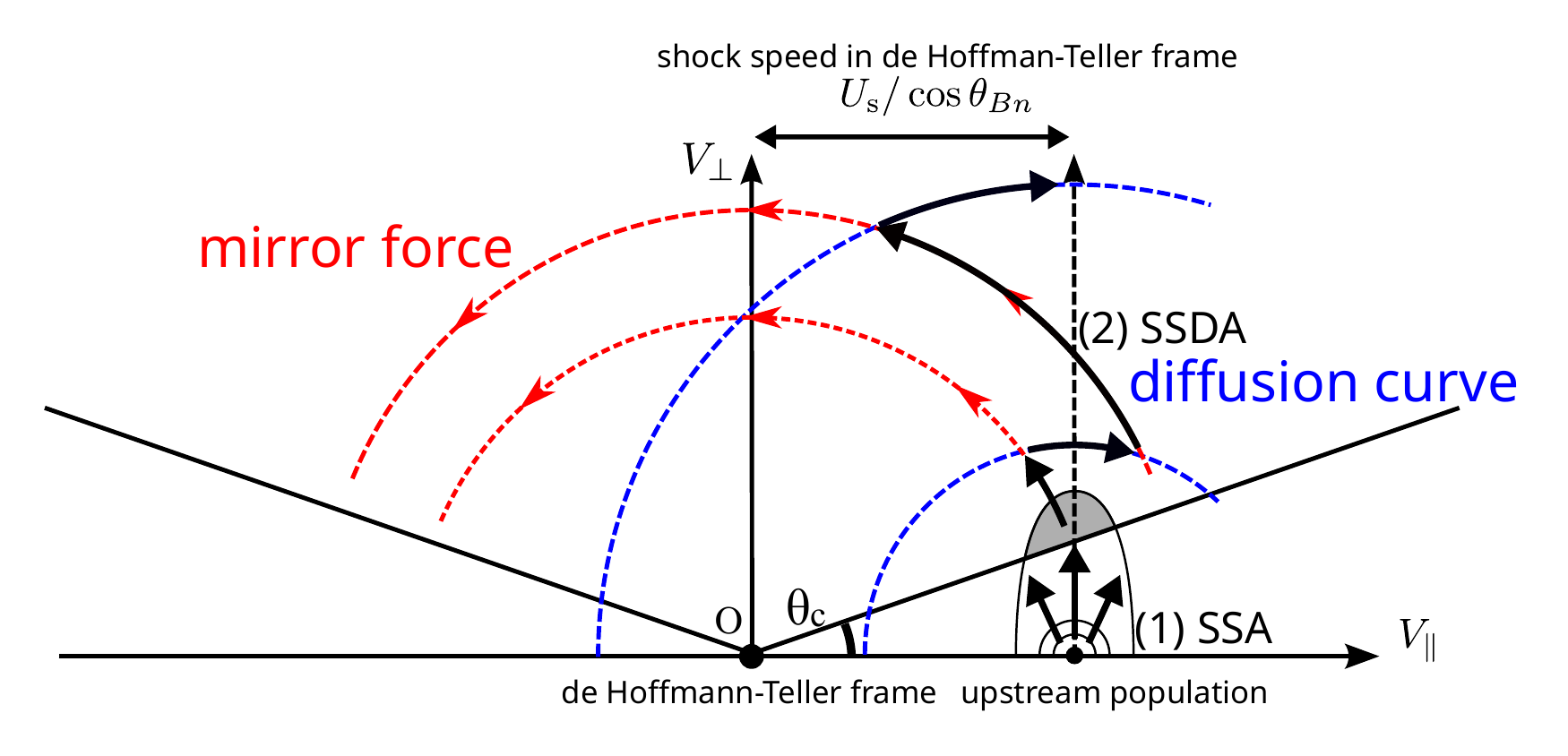}
    \caption{Schematic illustration of particle trajectory in velocity space expected by SSA followed by SSDA. It is nearly the same as \figref{fig:ssda} for SSDA, but now with the pre-acceleration phase by SSA. Some fraction of the cold upstream population energized by SSA will be accelerated further by SSDA.}
    \label{fig:ssa-ssda}
\end{figure}

SSA may yet play a role in electron injection even if the mechanism by itself is not effective enough in producing relativistic electrons. \citet{Amano2007} performed 1D PIC simulations of oblique shocks and demonstrated that the electrons pre-accelerated by SSA are accelerated further through the adiabatic mirror reflection by SDA. The reflection can be triggered by SSA even for low-energy electrons in the upstream that are initially well inside the loss cone. The two-step electron acceleration mechanism is realized for the following reasons: (1) SSA accelerates electrons predominantly perpendicular to the background magnetic field, (2) SSA is a process much faster than SDA, allowing the two processes to operate successively without interfering with each other. The second step will be replaced by SSDA if pitch-angle scattering occurs efficiently, which is illustrated in \figref{fig:ssa-ssda}. \citet{Matsumoto2017} found that SSA and SSDA both play the role of efficient electron acceleration in a fully 3D high Mach number shock. The scattering in the simulation by \citet{Matsumoto2017} is provided by the Weibel-generated turbulence, which is the subject of Section \ref{sec:5}.

To summarize, SSA associated with intense electrostatic waves generated by BI has been proven to be effective in 3D, albeit that the particle acceleration mechanism is somewhat different from the original one proposed based on the 1D assumption. It behaves rather like the ion SSA but with the macroscopic cross-shock potential being replaced by reflected-ion-driven turbulent electrostatic waves. Efficient SSA for electrons requires the condition given by \eqref{eq:unlimited-acc}. While the exact scaling laws with respect to the shock speed and the ion-to-electron mass ratio have not yet been understood, a single SSA may not be able to provide relativistic electrons by itself if the shock speed is of order $\Us/c \sim 0.01$ (typical of young SNR shocks). In this case, multiple SSAs might become important. Another point is that SSA may trigger subsequent particle acceleration by the adiabatic SDA or SSDA. If SSA is efficient enough to accelerate upstream cold electrons beyond the shock speed in HTF $\sim \Us/\cos\theta$, the two-step particle acceleration mechanism will be initiated at very high Mach number shocks, which would otherwise not be operative.

\clearpage
\section{Dynamics of Weibel-dominated Shocks} \label{sec:5}

\subsection{Weibel Instability} \label{sec:5-weibel}

The Weibel instability (WI) is an electromagnetic instability driven by an effective temperature anisotropy. The original paper by \citet{Weibel1959} analyzed the instability in the presence of a background magnetic field but found a purely growing mode in the unmagnetized plasma limit. Therefore, WI often refers to the instability in an unmagnetized plasma. As a result of instability, the initial free energy contained as temperature anisotropy is converted to the fluctuating magnetic field energy, which then scatters the particles to reduce the anisotropy.

The potential importance of WI in the context of relativistic unmagnetized collisionless shocks in gamma-ray bursts was pointed out by \citet{Medvedev1999b}. It has then been studied with PIC simulations extensively in subsequent work \citep[e.g.,][]{Silva2003,Hededal2004,Frederiksen2004,Medvedev2005,Kato2005a}. Since the shock transition layer is the place where two different plasmas (i.e., upstream and downstream plasmas) collide with each other, the overlap between the two streams provides an effective temperature anisotropy in momentum space. In other words, the effective thermal spread in the shock normal direction becomes much larger than in the transverse direction. The anisotropy drives WI, and the generated magnetic turbulence scatters the particles and leads to their thermalization. Such a shock in which the dissipation is mainly provided by the Weibel-generated turbulence has been reproduced with multidimensional PIC simulations \citep[e.g.,][]{Kato2007,Kato2008,Spitkovsky2008a}, which is often called a Weibel-mediated shock. Since the magnetic fluctuation amplitude should be strong enough to deflect the flow, a non-negligible fraction of the initial flow kinetic energy needs to be converted into the energy of magnetic turbulence. In gamma-ray bursts, WI is believed to be an important candidate for the magnetic-field generation that is inferred from observations. We note that sometimes anisotropy-driven WI is distinguished from a beam-driven filamentation instability \citep[e.g.][]{Bret2005a}. Although we will not make this distinction, there should be no confusion as the instability is always driven by the relatively cold reflected ion beam.

In the following, we will focus on WI in weakly magnetized non-relativistic shocks in electron-ion plasma, which we call Weibel-dominated shocks. We shall distinguish the magnetized Weibel-dominated shock from the Weibel-mediated shock in an unmagnetized plasma because they have slightly different characteristics. While the magnetic turbulence in the shock transition layer is substantial in both cases, the overall shock structure in the Weibel-dominated shock is still determined by the gyromotion of reflected ions in a background magnetic field. The situation is fundamentally different in the unmagnetized Weibel-mediated shock, in which back-streaming particles (i.e., those escaping upstream) will never come back to the shock unless they are scattered by turbulence. Consequently, there appears a precursor region of substantial spatial extent where a small fraction of back-streaming particles generate weak magnetic noise fluctuations \citep[e.g.,][]{Ruyer2016}. The situation is more or less similar to a magnetized parallel shock.

\subsection{Relation to Rippling Mode} \label{sec:5-rippling}

The finding that WI plays the dominant role in a magnetized non-relativistic shock transition layer came somewhat as a surprise \citep{Kato2010,Matsumoto2015}. In-situ spacecraft observations of shocks within the heliosphere indicate that WI is not a dominant mode, at least in the typical solar wind at 1 AU. Although the shock transition layer looks almost always highly turbulent, the magnetic fluctuation amplitude is normally comparable to the background magnetic field. Therefore, it is natural to conjecture that there is a regime transition from the classical magnetized shock to the more violent Weibel-dominated shock as the shock strength increases. While the shock speed is clearly a measure of the shock strength, it was not clear whether the transition is controlled by the shock speed normalized to the speed of light $\Us/c$ or by (\Alfven and sound) Mach numbers. The problem is that the shock speeds in typical PIC simulations of high Mach number shocks are very high $\Us/c \gtrsim 0.1$, much higher than those of typical young SNR shocks $\Us/c \sim 0.01$ or planetary bow shocks $\Us/c \sim 0.001$.

Conventionally, linear analysis of WI in the context of collisionless shocks has been performed by assuming that the plasma is unmagnetized. This is clearly a reasonable approximation if the magnetic field strength is sufficiently weak. The question is how weak it should be. Note that the unmagnetized plasma assumption corresponds to the limit of infinite \Alfven Mach number $\MA \rightarrow \infty$. It is known that the effective temperature anisotropy induced by the reflected ions gyrating around a background magnetic field can drive Alfven-ion-cyclotron (AIC) and mirror instabilities at a quasi-perpendicular shock with parameters typical of Earth's bow shock \citep{Winske1988}. Such a shock exhibits surface oscillations called rippling over the timescale of ion gyroperiod \citep[e.g.,][]{Johlander2016,Johlander2018}. While the exact relation between the rippling and the two competing instabilities has not been fully understood, it is well established that the effective ion temperature anisotropy is the free-energy source of the rippling mode. In other words, the same free energy appears to result in rippling oscillations or more violent Weibel-turbulence, depending on the strength of the shock.

To investigate the relation between WI and the rippling mode, one needs a model that can consistently reproduce both of them. \citet{Nishigai2021} investigated the reflected-ion-driven instability in a periodic system that mimics the shock transition layer. The system is homogeneous and consists of three distinct particle populations: The incoming electrons, ions, and the reflected ions. A gyrotropic ring velocity distribution function (VDF) in the plane perpendicular to the background magnetic field is used to represent the reflected ion component, while others are modeled with drifting Maxwellians. This model allows us to study the continuous transition from AIC and mirror instabilities at relatively weak shocks to WI at stronger shocks.

By combining 2D PIC simulations and linear analysis, \citet{Nishigai2021} found that WI is the mode on the same branch as the AIC wave. In other words, they both appear as the solutions to the single dispersion relation for circularly polarized electromagnetic waves of parallel propagation \citep{StixBook}:
\begin{align}
    D \left( \omega, k \right) = 1 - \frac{c^2 k^2}{\omega^2} +
    \pi \sum_{s} \frac{\omega_{ps}^2}{\omega^2}
    \int_{-\infty}^{+\infty} \int_{0}^{\infty}
    \frac{
        v_{\perp} k \frac{\partial F_{s}}{\partial v_{\parallel}} +
        \left( \omega - k v_{\parallel} \right) \frac{\partial F_{s}}{\partial v_{\perp}}
        }{\omega - k v_{\parallel} - \Omega_{s}}
    v_{\perp}^2 d v_{\perp} d v_{\parallel} = 0.
    \label{eq:parallel-dispersion}
\end{align}
Numerical solutions obtained from \eqref{eq:parallel-dispersion} confirm that they smoothly transition from one another. It may be interesting to mention that the approximate analytic expression of the maximum growth rate for WI obtained in the short wavelength and high \Alfven Mach number limit
\begin{align}
    \frac{\gammax}{\wci} \approx
    \left( \frac{n_r}{2 n_0} \right)^{1/2} \MA
    \label{eq:weibel-growthrate}
\end{align}
is formally equivalent to that for the AIC instability obtained in the limit of $\gamma/\wci \ll 1$ \citep{Nishigai2021}. Here, $n_r/n_0$ is the reflected ion density relative to the background. We should note that the assumption of parallel propagation is adequate as the perturbations seen in PIC simulations for high Mach numbers are dominated by the parallel propagating modes with no evidence for the obliquely propagating mirror-mode waves. The simulation results showed that \Alfvenic perturbations characteristics of the AIC instability are seen if the growth rate is relatively small $\gamma/\wci \lesssim 1$. On the other hand, as the instability growth rate increases, the evolution of the system becomes more Weibel-like in the sense that the initially growing short-wavelength current filaments merge with each other to generate long-wavelength modes over a timescale less than $\wci^{-1}$. This indicates that the instability will behave as Weibel-like if the growth rate is much larger than the ion gyrofrequency ($\gamma/\wci \gg 1$).

\begin{figure}[tbp]
    \centering
    \includegraphics[width=0.60\textwidth]{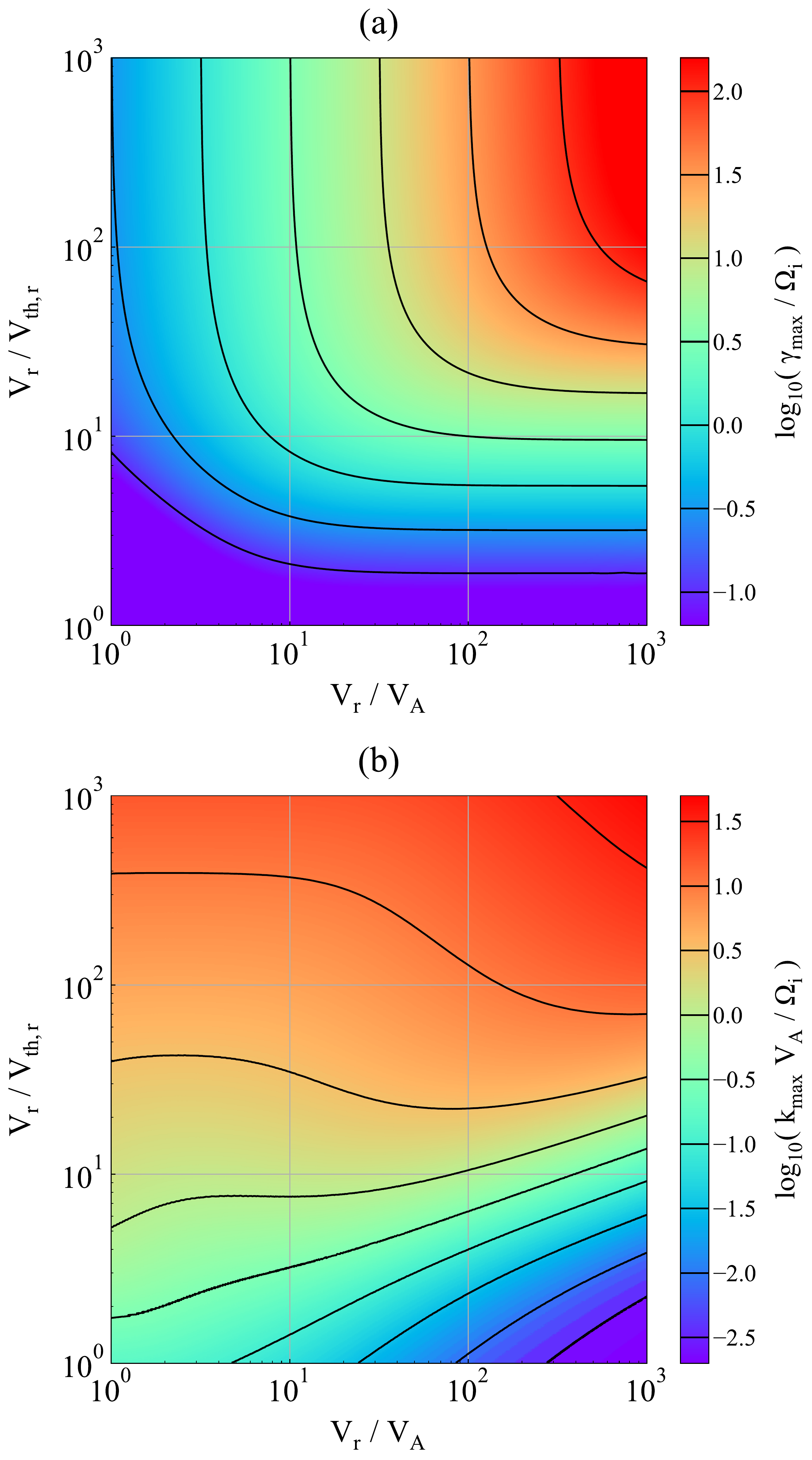}
    \caption{Summary of fully kinetic linear analysis obtained with the reflected ion component modeled by a magnetized ring distribution: (a) maximum growth rate, (b) wavenumber at maximum growth. A reflected ion density of $n_r/n_0 = 0.2$ is used. The horizontal and vertical axes are proxies for \Alfven and sound Mach numbers, respectively. This figure is reproduced from \citet{Nishigai2021}.}
    \label{fig:nishigai2021}
\end{figure}

\figref{fig:nishigai2021} shows the survey of linear analysis as a function of $V_{r}/V_\textrm{th,r}$ and $V_{r}/\VA$ where $V_r$ is the ring velocity, $V_\textrm{th,r}$ is the ring thermal velocity, and $\VA$ is the \Alfven velocity, respectively. Since the ring velocity is essentially the same as the shock speed (i.e., $V_r \sim \Us$), the parameters for the vertical and horizontal axes are understood as proxies for sound and \Alfven Mach numbers, respectively. The panels (a) and (b) show the maximum growth rate $\gammax/\wci$ and the wavenumber at the maximum growth $k_\textrm{max} \VA/\wci$, respectively. It is important to note that the result is not sensitive to $m_i/m_e$ and $\wpe/\wce \propto c/\VA$, indicating that $\Us/c$ is not a relevant parameter regulating the instability characteristics (as long as $\Us/c \ll 1$). The transition from $\gammax/\wci \lesssim 1$ to $\gammax/\wci \gg 1$ requires that both the \Alfven and sound Mach numbers need to be sufficiently high. More specifically, the transition appears to occur at Mach numbers of around $\sim 20{\rm -}40$. Therefore, planetary bow shocks and interplanetary shocks in the heliosphere will unlikely be in the Weibel-dominated regime except for unusual solar wind conditions. Note that high \Alfven Mach number Earth's bow shocks reported by \citet{Sundberg2017a} and \citet{Madanian2020} were in high-$\beta$ solar wind conditions (i.e., relatively low sound Mach number) and will not be dominated by WI. Possible exceptions might be found at high Mach number planetary bow shocks in the outer heliosphere where the adiabatic cooling of the solar wind may increase both the sound and \Alfven Mach numbers \citep[e.g.,][]{Masters2013}. In contrast, young SNR shocks with very high Mach numbers will likely be Weibel-dominated shocks involving strong magnetic turbulence $\delta B/B_0 \gg 1$. We note that the magnetic field amplification suggested by astrophysical observations \citep[e.g.,][]{Uchiyama2007} has often been interpreted as a result of MHD turbulence \citep{Giacalone2007a,Inoue2009} or a non-resonant cosmic-ray-driven instability at a quasi-parallel shock \citep{Winske1984a,Lucek2000a,Bell2004}. The Weibel-turbulence provides an alternative to these models, which may be most efficient at quasi-perpendicular shocks and is consistent with in-situ measurements of Saturn's bow shock \citep{Bohdan2021}.

\subsection{Spontaneous Magnetic Reconnection} \label{sec:5-reconnection}

\begin{figure}[tbp]
    \centering
    \includegraphics[width=1.00\textwidth]{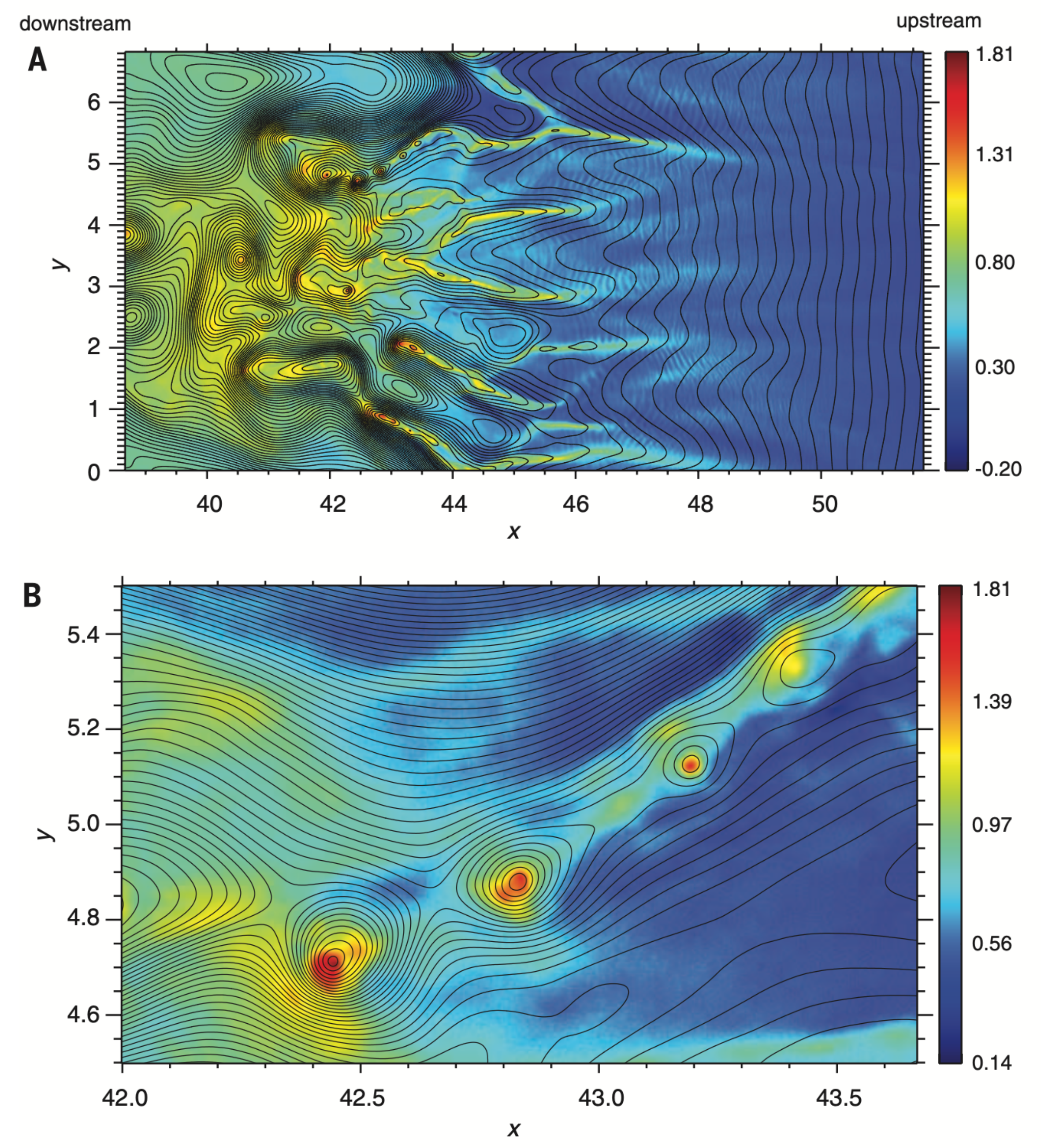}
    \caption{Magnetic reconnection found in the 2D simulation by \citet{Matsumoto2015}. The black lines show the magnetic field lines, and the color represents the density. The overview of the entire shock transition layer is shown in the panel (A), while the enlarged view around the actively reconnecting current sheet is shown in the panel (B). This figure is reproduced from \citet{Matsumoto2015}.}
    \label{fig:matsumoto2015a}
\end{figure}

The interesting consequence of the Weibel-dominated shocks first reported by \citet{Matsumoto2015} is that a lot of current sheets are produced in their nonlinear development within the shock transition layer, which eventually break up spontaneously via magnetic reconnection. \figref{fig:matsumoto2015a} is reproduced from the 2D PIC simulation result by \citet{Matsumoto2015}, which shows the magnetic field lines (the solid black lines) and the density (color) in the shock transition layer. A series of magnetic islands with enhanced density clearly indicate ongoing magnetic reconnection activity.

Since magnetic reconnection in a 2D simulation requires magnetic field lines with opposite polarities contained in the simulation plane, the amplification of the in-plane magnetic field to a level much larger than the background $\delta B/B \gg 1$ is needed. Naively speaking, such an in-plane magnetic field will not be generated by WI because it predicts the growth of the magnetic field component perpendicular to both the beam direction and the wavenumber vector. If the background magnetic field is absent or in the out-of-plane direction, it is only the out-of-plane magnetic field that can be generated by WI. Consequently, magnetic reconnection is prohibited in these cases. Therefore, the presence of a finite in-plane background magnetic field is a necessary condition for spontaneous magnetic reconnection. Note that the situation is opposite to SSA, for which the out-of-plane configuration provides the most efficient particle acceleration (see, Section \ref{sec:4}). It was indeed the motivation of \citet{Bohdan2017} who performed 2D simulations of purely perpendicular shocks with background magnetic fields inclined by $45 \degr$ from the plane of simulations, in trying to incorporate the two important ingredients in a 2D simulation domain.

\citet{Matsumoto2015} suggested that an effective ion anisotropy in the out-of-plane direction induced by the gyromotion of reflected ions in the presence of a finite in-plane magnetic field can also provide the free energy source. Such anisotropy will produce magnetic perturbations in the shock normal direction that might lead to magnetic reconnection. However, we have recently confirmed that the in-plane magnetic field can be generated even without an out-of-plane anisotropy with 2D PIC simulations in a periodic simulation box. A possible explanation for this is field-line stretching by Weibel-generated electron flow velocity fluctuations. Since we consider timescale on the order of $\wci^{-1}$, ions are essentially unmagnetized. However, the magnetic field lines will be frozen-in to the electron flow if $\gammax/\wce = (m_e/m_i) \gammax/\wci < 1$ such that the electrons behave magnetized. Therefore, if the in-plane electron velocity fluctuations are generated, the in-plane magnetic field can be amplified by a mechanism similar to the MHD dynamo action. Indeed, the out-of-plane magnetic perturbations driven by WI will induce in-plane currents that may be carried by electrons. We have also performed a preliminary linear analysis showing that the reflected-ion-driven WI in the presence of a magnetized electron component can generate an in-plane magnetic field component. The presence of a background magnetic field and a sufficiently large mass ratio (allowing the decoupling between the ion and electron dynamics) are two important factors leading to the generation of in-plane magnetic turbulence and, ultimately, magnetic reconnection. A more detailed discussion will be presented elsewhere in a future publication.

Understanding the in-plane magnetic field generation mechanism is obviously not enough to predict the condition for magnetic reconnection within the shock transition layer. \citet{Bohdan2020a} clearly showed that the magnetic reconnection activity increases with increasing the mass ratio and the \Alfven Mach number. Naively speaking, magnetic reconnection may be initiated when the generated magnetic field becomes much larger than the background magnetic field $\delta B/B_0 \gg 1$. \citet{Bohdan2020a} argued that magnetic reconnection will be activated if the growth time of WI ($\gammax^{-1}$) is sufficiently short (say, by a factor of 1/10) compared to the plasma convection timescale over the shock transition layer ($\wci^{-1}$). If we use the analytic estimate of the growth rate \eqref{eq:weibel-growthrate} and adopt the same argument by \citet{Bohdan2020a}, we obtain the following condition.
\begin{align}
    \MA \gtrsim 10 \left( \frac{n_r}{2 n_0} \right)^{-1/2}
    \approx 32 \left( \frac{n_r/n_0}{0.2} \right)^{-1/2}.
    \label{eq:mrx-condition}
\end{align}
We note that this is essentially the same as the condition $\gammax/\wci \gtrsim 10$, at which the shock will behave as Weibel-dominated (see, Section \ref{sec:5-rippling}). This result suggests that the magnetic field amplification efficiency will be independent of the mass ratio, as was also shown clearly in \citet{Bohdan2021}. It is, nevertheless, important to point out that the dynamo-like magnetic field amplification suggested above requires a scale separation between the ion and electron dynamics. This indicates that the mass ratio might potentially play a role in regulating the magnetic reconnection activity.

While it is possible to anticipate that the spontaneously generated current sheets may eventually be subject to the tearing-mode instability, there may be other kinds of instabilities that compete with the tearing mode. For instance, the drift-kink, lower-hybrid-drift, and Kelvin-Helmholtz instabilities have been considered as important instabilities in a current sheet \citep[e.g.,][]{Davidson1977,Daughton1998a,Shinohara2001}. It is, however, quite possible that those models considered for the Harris current sheet might not simply apply to a non-Harris current sheet developed in the shock transition layer. It is also important to mention that the magnetic field generated by WI in an unmagnetized plasma eventually dissipates via the kink instability rather than magnetic reconnection in its nonlinear phase \citep{Ruyer2018}. We suggest that the absence of the dynamo-like magnetic field amplification mechanism may exclude the dissipation via magnetic reconnection in an unmagnetized plasma. In any case, it is interesting to investigate whether a finite background magnetic field regulates the ultimate fate of current sheets in the shock transition layer. Fully 3D PIC simulations are, however, necessary to address this issue. The 3D simulation result by \citet{Matsumoto2017} appears less efficient in terms of magnetic reconnection activity, which might be because the Mach numbers of the shock ($\MA \sim 20.8$ and $\MS \sim 22.8$) are slightly smaller than the predicted threshold condition \eqref{eq:mrx-condition}.

\subsection{Particle Acceleration} \label{sec:5-particle-acc}

The spontaneous magnetic reconnection within the shock transition layer indicates that there is a new energy dissipation channel that becomes activated only at high Mach numbers shocks. In other words, the upstream flow kinetic energy once converted into the turbulent magnetic energy, will ultimately dissipate via magnetic reconnection, which may have important implications for particle acceleration. Magnetic reconnection has long been believed as one of the most probable mechanisms for high-energy particle acceleration in heliospheric and astrophysical plasmas. Although it is capable of accelerating both ions and electrons, it may accelerate electrons more efficiently than a collisionless shock does. As we have discussed in Section \ref{sec:2-electron-injection}, the collisionless shock may not necessarily be an efficient electron accelerator. In contrast, numerous observational evidence suggests that magnetic reconnection produces nonthermal electrons relatively efficiently \citep{Oieroset2002,Lin2003,Imada2011}. This suggests that the activation of magnetic reconnection in a shock potentially enhances the electron acceleration efficiency at high Mach number shocks.

\begin{figure}[tbp]
    \centering
    \includegraphics[width=1.00\textwidth]{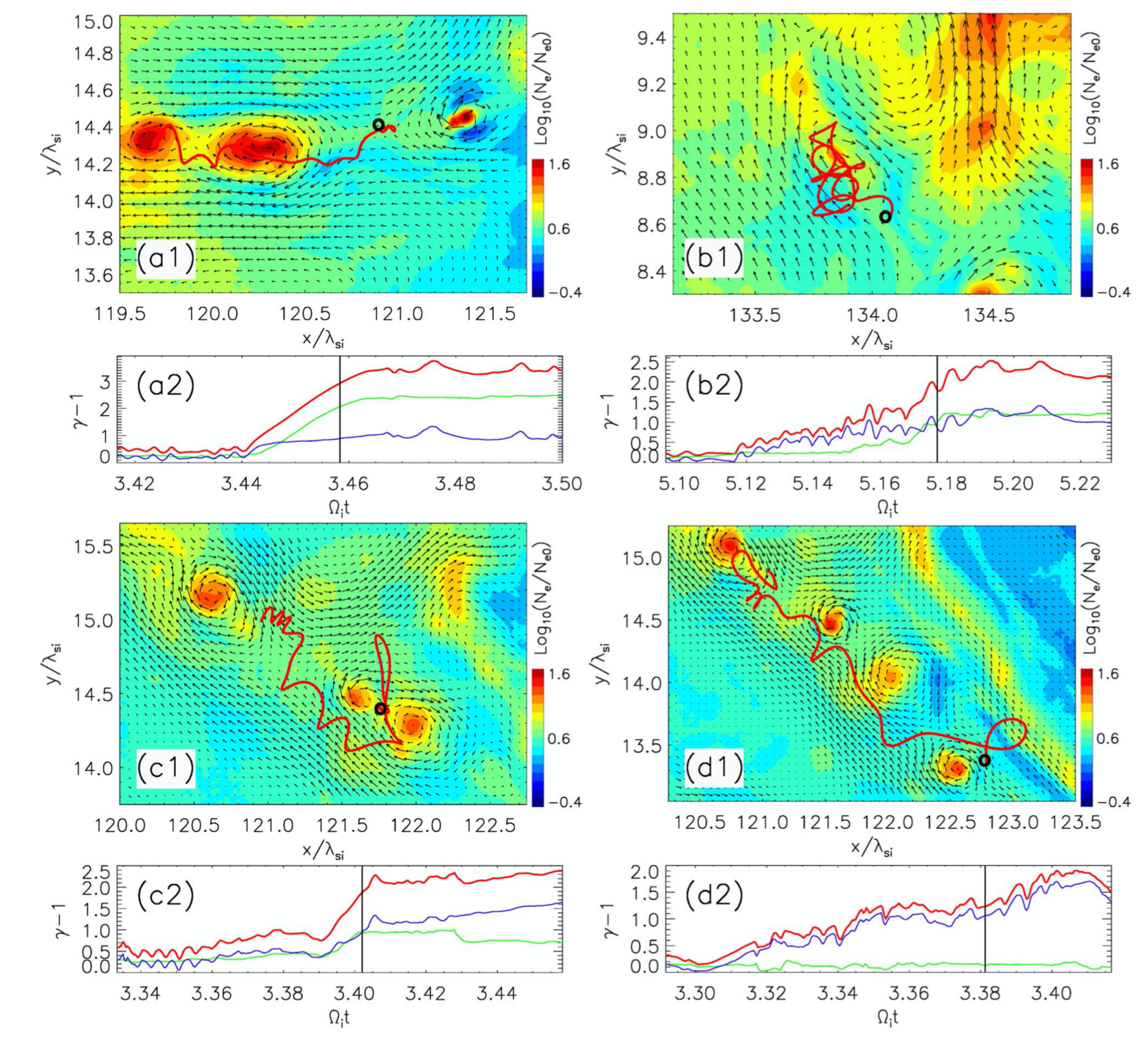}
    \caption{Trajectories of accelerated particles associated with magnetic reconnection activity in the simulation by \citet{Bohdan2020a}. Each panel shows the density in color and the magnetic field vector with arrows. The particle trajectories in the $x-y$ plane, as well as their energy histories, are shown. The red, green, and blue lines in the energy history show the total, parallel, and perpendicular energies, respectively. This figure is reproduced from \citet{Bohdan2020a}.}
    \label{fig:bohdan2020a}
\end{figure}

\citet{Matsumoto2015} first reported that nonthermal electron acceleration takes place associated with spontaneous magnetic reconnection. Various mechanisms have been considered for particle acceleration during magnetic reconnection, including (1) DC electric field acceleration around the X-line followed by gradient-B and curvature-B drifts in the reconnection jet-plasmoid interaction region \citep{Speiser1965a,Hoshino2001}, (2) first-order Fermi acceleration in contracting magnetic islands \citep{Drake2006} (3) island surfing acceleration \citep{Oka2010b}, (4) stochastic acceleration in turbulent multiple reconnecting current sheets \citep{Hoshino2012,Drake2013}. See, \citet{Oka2010a} for more details on various particle acceleration mechanisms. A detailed analysis performed by \citet{Bohdan2020a} indeed found that a variety of particle acceleration mechanisms are operating in a high Mach number shock. \figref{fig:bohdan2020a} shows examples of energetic electron trajectories during interaction with reconnecting current sheets. \citet{Bohdan2020a} identified particle acceleration mechanisms in each panel as follows: (a) DC electric field acceleration at the X-line, (b) island surfing acceleration, (c) first-order Fermi acceleration in merging magnetic islands, (d) stochastic second-order Fermi-like acceleration. It may be surprising that the particle acceleration mechanisms seen within the shock transition layer are markedly similar to those found in simulations of magnetic reconnection starting from the classical Harris current sheet equilibrium.

It is nevertheless not easy to quantitatively evaluate the relative contribution of magnetic reconnection to the final nonthermal electron spectrum. \citet{Bohdan2020a,Bohdan2020b} differentiated contributions of different electron acceleration and heating mechanisms in 2D PIC simulations obtained with the in-plane configuration. We should, however, keep in mind that magnetic reconnection is active only in the in-plane configuration, whereas shock surfing acceleration (SSA), which is likely to be efficient in the same parameter regime, is most efficient in the out-of-plane configuration (see Section \ref{sec:4}). The strong dependence of the dynamics on the background magnetic field orientation suggests that it may not be easy to draw a final conclusion from 2D simulations alone.

As we already mentioned, the 3D simulation by \citet{Matsumoto2017} found that the particle acceleration occurs in a combination of SSA and SSDA. In contrast, we have not yet been able to identify a signature of particle acceleration by magnetic reconnection in the fully 3D shock transition layer. It is possible that the Weibel-generated turbulence in the simulation is not strong enough for driving efficient magnetic reconnection because of relatively lower Mach numbers. If this is the case, we will be able to find active magnetic reconnection and associated particle acceleration in a higher Mach number shock. Alternatively, the 3D effect may introduce a non-negligible impact on the dynamics of magnetic reconnection. For instance, the drift-kink instability may have a growth rate larger than the tearing-mode instability, in particular, with a lower mass ratio \citep[e.g.,][]{Daughton1999b}. Consequently, the efficiency of particle acceleration may also be substantially modified \citep{Zenitani2007a}. The SSA-SSDA two-step particle acceleration mechanism found in the particular simulation by \citet{Matsumoto2017} may continue to operate with an added effect of magnetic reconnection at even higher Mach numbers. However, if magnetic reconnection fully dominates the shock dynamics, the concept of the gradient-B drift itself may become inappropriate in such a highly turbulent shock structure. It might thus be possible that SSDA based on the energy gain via the gradient-B drift does not survive in reality. In summary, the Weibel-dominated shock is a relatively new regime of collisionless shocks, and a lot of questions remain yet unanswered. It is clear that future investigation of the dynamics with fully 3D PIC simulations is still necessary.


\clearpage
\section{Summary and Prospect} \label{sec:6}

In this review, we have discussed the non-thermal electron acceleration processes in quasi-perpendicular collisionless shocks. It is not our intention to present a fully comprehensive review on the broad subject. Instead, we have focused on more specific topics that have been obtained mostly by the authors in the last several years.

In Section \ref{sec:3}, we have presented an up-to-date understanding of stochastic shock drift acceleration (SSDA), which we think, at present, is the most probable mechanism for electron injection into diffusive shock acceleration (DSA). Kinetic simulations suggest that energetic particle trajectories appear to be consistent with the theoretical prediction. In-situ spacecraft observations at Earth's bow shock have also found solid evidence for the theory. The theoretical model is capable of predicting the accelerated particle spectrum for a broad energy range, both below and above the injection threshold energy. In general, a high \Alfven Mach number shock defined in the de Hoffmann-Teller frame (HTF) is favorable for electron injection. The theory predicts that, if certain conditions are satisfied, a steep low-energy spectrum produced by SSDA will be connected to the harder DSA power-law index at high energy.

Understanding the wave generation mechanisms and the resulting particle scattering efficiency is the key to more sophisticated and quantitative modeling of electron injection at shocks. Existing theoretical models need to be tested with in-situ observations and fully kinetic simulations. It may also be important to investigate the heating efficiency of thermal electrons as it may provide a seed population for SSDA. Given this information, it will be possible to construct the entire electron energy spectrum from thermal to ultra-relativistic energies, which may be compared with radio and X-ray synchrotron radiation from astrophysical sources.

In Section \ref{sec:4} and \ref{sec:5}, we have mainly discussed recent advances on the dynamics of very high Mach number shocks obtained with 2D and 3D PIC simulations. Section \ref{sec:4} is devoted to discussion of shock surfing acceleration (SSA), which occurs associated with high-frequency large-amplitude electrostatic waves generated by the Buneman instability in the leading edge of the shock transition layer. It was proposed originally based on the assumption that the electrostatic potential is homogeneous in the direction transverse to the wave propagation direction. However, it is not realistic even in the linear phase of instability as many modes propagating obliquely with respect to the reflected ion beam will be unstable. Interestingly, 2D and 3D simulations found electron acceleration occurs in a slightly different manner compared to 1D even with highly inhomogeneous electrostatic potential structures. The particle acceleration takes place in association with a specular reflection by the turbulent electrostatic waves, which act like a wall moving with the driving reflected ions. We suggest that the energy gain by a single reflection is not necessarily sufficient for electron injection with realistic shock speeds at young SNRs. However, the reflection, in principle, may occur multiple times at a realistic shock, which will produce even higher energy electrons. Alternatively, even if SSA itself does not provide sufficient energy gain for injection, it may act as a pre-acceleration mechanism for further acceleration through SSDA.

In Section \ref{sec:5}, we have argued that the Weibel instability will be unstable in high Mach number young SNR shocks. Such a shock will likely host both the Buneman and Weibel instabilities simultaneously in different regions of the shock transition layer, although the exact instability conditions may not necessarily be the same. Intense Weibel-generated turbulence will be able to scatter electrons accelerated by SSA. Therefore, the pre-accelerated population may preferentially be subjected to further particle acceleration by SSDA. Another interesting consequence of the Weibel-dominated shock is that magnetic reconnection may occur spontaneously because the turbulence produces a lot of current sheets within the shock transition layer. In other words, the upstream kinetic energy is converted into magnetic energy through the Weibel instability, which eventually dissipates via magnetic reconnection. The spontaneous magnetic reconnection opens up a new way of accelerating particles at a shock. A variety of particle acceleration mechanisms discussed in the magnetic reconnection literature have been identified within the shock transition layer. While it is reasonable to assume that magnetic reconnection is activated only at sufficiently high Mach number shocks, the exact condition has not yet been understood. Furthermore, the current sheets in 3D may break up with instabilities other than the tearing-mode instability. This indicates a need for fully 3D PIC simulations to understand the fate of spontaneously generated current sheets in a real shock. The question is whether the SSDA theory constructed for the well-established classical magnetized shock may or may not be applicable to violent Weibel-dominated shocks in which magnetic dissipation processes are ongoing.

\backmatter


\bmhead{Acknowledgments}
The work of T.~A. was supported by JSPS KAKENHI Grant Nos.~17H02966 and 17H06140. The work of J.~N. has been supported by Narodowe Centrum Nauki through research project No. 2019/33/B/ST9/02569. The authors acknowledge the support by the International Teams program of International Space Science Institute (ISSI).

\bmhead{Conflict of interest}
On behalf of all authors, the corresponding author states that there is no conflict of interest.


\bibliography{reference}



\end{document}